\documentclass[12pt,preprint]{aastex}
\usepackage{lscape}
\usepackage{epstopdf}
\usepackage{graphicx,epsfig,emulateapj5}
\usepackage{amsmath}
\usepackage{natbib}
\newcommand{\etal}{{et al.~}}

\newcommand{\bq}{\begin{equation}}
\newcommand{\eq}{\end{equation}}

\newcommand{\ee}[1]{\mbox{${} \times 10^{#1}$}}

\newcommand{\h}{\mbox{$^h$}}

\newcommand\cmv{\mbox{cm$^{-3}$}}

\newcommand{\spitzer}{\mbox{\it Spitzer}}

\newcommand{\lsun}{\mbox{L$_\odot$}}
\newcommand{\msun}{\mbox{M$_\odot$}}

\newcommand{\mean}[1]{\mbox{$\langle#1\rangle$}} 

\newcommand{\hcop}{HCO$^+$}

\def\gtsim{\lower.5ex\hbox{$\buSildrel > \over\sim$}}
\def\ltsim{\lower.5ex\hbox{$\buildrel < \over\sim$}}
\def\arcsec{^{\prime\prime}}

\def\apjl{ApJL}
\def\apj{ApJ}
\def\apjs{ApJS}
\def\mnras{MNRAS}
\def\araa{ARAA}
\def\aj{AJ}
\def\aap{A\&A}

\def\pasp{PASP}
\def\h1{H{\sc \! i}}

\newcommand{\msunyr}{\mbox{M$_\odot$ yr$^{-1}$}}

\slugcomment{Accepted by the Astrophysical Journal}
\shortauthors{Heiderman \etal}
\shorttitle{Galactic Star Formation Rate and Gas Surface Density Relations}

\begin{document}

\title{The Star Formation Rate and
 Gas Surface Density Relation in the Milky Way: Implications for
 Extragalactic Studies}

\author{Amanda Heiderman\altaffilmark{1}, Neal J. Evans
 II\altaffilmark{1}, Lori E. Allen\altaffilmark{2}, Tracy
 Huard\altaffilmark{3}, Mark Heyer\altaffilmark{4}}

\altaffiltext{1}{Department of Astronomy, University of Texas at
 Austin, 1 University Station C1400, Austin, TX
 78712--0259, USA; alh@astro.as.utexas.edu, nje@astro.as.utexas.edu}
\altaffiltext{2}{National Optical Astronomy Observatory, 950 North Cherry Avenue, Tucson, AZ 85719, USA; lallen@noao.edu}
\altaffiltext{3}{Department of Astronomy, University of Maryland,
 College Park, MD 20742, USA; thuard@astro.umd.edu}
\altaffiltext{4}{Department of Astronomy, University of Massachusetts,
 Amherst, MA 01003--9305, USA; heyer@astro.umass.edu}

\begin{abstract}
We investigate the relation between star formation rate (SFR) and gas
surface densities in Galactic star forming regions using a sample of
young stellar objects (YSOs) and massive dense clumps.  Our YSO sample
consists of objects located in 20 large molecular clouds from the
\spitzer \ cores to disks (c2d) and Gould's Belt (GB) surveys.  These
data allow us to probe the regime of low--mass star formation
essentially invisible to tracers of high--mass star formation used to
establish extragalactic SFR-gas relations.  We estimate the gas
surface density ($\Sigma_{\rm gas}$) from extinction ($A_{V}$) maps
and YSO SFR surface densities ($\Sigma_{\rm SFR}$) from the number of
YSOs, assuming a mean mass and lifetime. We also divide the clouds
into evenly spaced contour levels of $A_{V}$, counting only Class I
and Flat SED YSOs, which have not yet migrated from their birthplace.
For a sample of massive star forming clumps, we derive SFRs from the
total infrared luminosity and use HCN gas maps to estimate gas surface
densities.  We find that c2d and GB clouds lie above the extragalactic
SFR-gas relations (e.g., Kennicutt--Schmidt Law) by factors up to 17.
Cloud regions with high $\Sigma_{\rm gas}$ lie above extragalactic
relations up to a factor of 54 and overlap with high--mass star
forming regions.  We use $^{12}$CO and $^{13}$CO gas maps of the
Perseus and Ophiuchus clouds from the COMPLETE survey to estimate gas
surface densities and compare to measurements from $A_{V}$ maps. We
find that $^{13}$CO, with the standard conversions to total gas,
underestimates the $A_{V}$-based mass by factors of $\sim$4--5.
$^{12}$CO may underestimate the total gas mass at $\Sigma_{\rm
gas}\gtrsim$200 $M_{\sun} \ \rm pc^{-2}$ by $\gtrsim$30\%; however,
this small difference in mass estimates does not explain the large
discrepancy between Galactic and extragalactic relations.  We find
evidence for a threshold of star formation ($\Sigma_{\rm th}$) at
129$\pm$14 $M_{\sun} \ \rm pc^{-2}$. At $\Sigma_{\rm gas} >
\Sigma_{\rm th}$, the Galactic SFR-gas relation is linear.  A possible
reason for the difference between Galactic and extragalactic relations
is that much of $\Sigma_{\rm gas}$ is below $\Sigma_{\rm th}$ in
extragalactic studies, which detect all the CO-emitting gas. If the
Kennicutt--Schmidt relation ($\Sigma_{\rm SFR} \propto \Sigma_{\rm
gas}^{1.4}$) and a linear relation between dense gas and star
formation is assumed, the fraction of dense star forming gas ($f_{\rm
dense}$) increases as $\sim \Sigma_{\rm gas}^{0.4}$.  When
$\Sigma_{\rm gas}$ reaches $\sim$300$\Sigma_{\rm th}$, the fraction of
dense gas is $\sim$1, creating a maximal starburst.

\end{abstract}

\keywords{stars: formation --- infrared: stars ---ISM: dust --- ISM:
 clouds --- galaxies: ISM}

\section{Introduction}\label{intro}

Understanding how physical processes in the interstellar medium (ISM)
control star formation is an important  prerequisite to
understanding galaxy evolution.  A robust measurement of the relation
between the star formation rate surface density ($\Sigma_{\rm SFR}$)
and the surface density of cold gas ($\Sigma_{\rm gas}$) is of vital
importance for input into theoretical models of galaxy evolution.

The idea that there should be a relation between the density of star
formation and gas density was first proposed by \citet{schmidt59}.  Schmidt
investigated this relation, now known as the ``Schmidt law", assuming
that it should be in the form of a power law and suggested that the
density of star formation was proportional to gas density squared.
\citet{k98} measured the global or disk--averaged  Schmidt law in a
sample of spiral and starburst galaxies using the projected star
formation and gas surface densities ($\Sigma_{\rm gas}$) in the form:

\bq
\Sigma_{\rm SFR} \propto \Sigma_{\rm gas}^{N},
\eq

where N is the power law index.  The global SFR and $\Sigma_{\rm gas}$
measurements for the sample of galaxies in \citet{k98} were fitted to a
power law with  $N=$1.4, which is known as in the ``Kennicutt--Schmidt
law" :

\bq\label{k98eq}
\Sigma_{\rm SFR} = (2.5\pm 0.7) \times 10^{-4}
\left (\frac{\Sigma_{\rm gas}}{1 M_{\sun} \rm \ pc^{-2}}
\right)^{1.4\pm 0.15}
 (M_{\sun} \ \rm yr^{-1} \ \rm kpc^{-2}).
\eq

 Since it is only an assumption that there is {\it only one relation}
that regulates how gas is forming stars, we refrain from calling this
a ``law" and instead refer to it as a SFR--gas relation, or as the
Kennicutt-Schmidt relation when referring to equation 2 specifically.
Several authors \citep{larson92,elmegreen94,wong02,krumtan07} argue
that there is a simple explanation for the Kennicutt--Schmidt
relation: if the SFR is proportional to the gas mass divided by the
time it takes to convert the gas into stars and if we take this
timescale to be the free--fall time, then $t_{ff}\propto \rho_{\rm
gas}^{-0.5}$ and $\dot{\rho}_{\rm SFR}\propto \rho_{\rm
gas}^{1.5}$.  Taking the scale height to be constant, $\rho
\propto\Sigma$, and this in turn gives the Kennicutt-Schmidt relation
(to the extent that $1.4\pm 0.15 = 1.5$).  A variety of observational
methods have been used to investigate this relation in different types
of galaxies and on different scales.

 There have been many observational studies of SFR--gas relations on
 either global scales \citep{kennicutt89, k98} or using either radial
 \citep{martin01,wong02,boissier03,heyer04,komugi05,schuster07} or
 point--by--point measurements \citep{kuno95,zfw01} that find
 values of  $N$ ranging from 1--2.  Recently, there have been studies that measure star formation and gas content of galaxies on spatially resolved scales of $\sim$0.2-2 kpc.
These studies have found power law indices of $N\approx0.8-1.6$
\citep{kennicutt07,thilker07,bigiel08, braun09,blanc09,verley10}.  The
study by \citet{bigiel08} used a sample of 18 nearby galaxies to
derive a spatially resolved relation on $\sim$750 pc scales.  They
found a linear relation between $\Sigma_{\rm SFR}$ and the molecular
gas surface density over a range of 3--50 $M_{\sun} \rm pc^{-2}$:

\bq\label{b08eq}
\Sigma_{\rm SFR} =  10^{-2.1\pm 0.2} \left( \frac{\Sigma_{\rm H_{2}}}{10 M_{\sun} \rm \ pc^{-2}} \right)^{1.0\pm 0.2} (M_{\sun} \ \rm yr^{-1} \ \rm kpc^{-2}).
\eq

Other spatially resolved studies were based on measurements done in a
single galaxy on scales of $\sim$100--500 pc: M51 \citep{kennicutt07,
blanc09}, NGC 7331 \citep{thilker07}, M31 \citep{braun09}, and M33
\citep{verley10}. Since the global study of \citet{k98} and spatially
resolved study of \citet{bigiel08} obtain results for large samples of
galaxies, we use these studies as a baseline for comparison to our
work. It is evident that sensitivity of $N$ to systematic variations
in methodology (e.g., data spatial resolution, SFR tracers, power law
fitting method) accounts for some of the differences in the derived
star formation power law index, but the underlying physical reasons
for the variations in the SFR--gas relations remain an open,
challenging question.

\citet{krumholz09} revisited the SFR--gas
relation, considering the dependence on atomic and molecular
components of $\Sigma_{\rm gas}$, metallicity, and clumping of the
gas.  Their analysis produces a SF--gas relation that rises steeply
at low $\Sigma_{\rm gas}$, where the gas is mostly atomic, is nearly
linear in the regime where normal spiral galaxies are found
\citep{kennicutt07,bigiel08,blanc09}, and increases superlinearly
above 85 $M_{\sun}\ \rm pc^{-2}$. Measurements made in these
studies, however, are limited to hundred parsec scales or more and are
not directly comparable to the size of individual molecular clouds or
dense clumps where stars form.  While these studies have all looked at
the extragalactic SFR--gas relation, there has been little work until
recently investigating this relation locally in the Milky Way.

Surveys of nearby molecular clouds in the Milky Way using $\spitzer$
imaging have provided large statistical samples of young stellar
object (YSO) candidates (e.g., Allen \etal 2010, in preparation,
\citet{evans09}, \citet{for09}, \citet{rebull10}).  These surveys have
allowed us to directly count the number of low--mass stars that are
forming and estimate SFRs.  These data also allow us to trace the
low--mass star formation regime essentially invisible to tracers, such
as emission in H$\alpha$, ultraviolet, far--infrared (FIR), and singly
ionized oxygen, used to establish extragalactic SFR-gas
relations. Since these tracers only probe the rate at which massive
stars form, a stellar Initial Mass Function (IMF), extrapolating down
to low stellar masses, must be assumed to obtain a SFR. Thus, these
SFR estimates are very sensitive to the IMF slope and distribution on
the low--mass end.  

\citet{evans09} compared extragalactic observed SFR--gas relations to
total molecular cloud measurements from the $\spitzer$ c2d survey.
They found that Galactic clouds lie above the SFR--gas relations
predicted by extragalactic work \citep{bigiel08, k98} and lie slightly
above the extrapolated relation from a study of massive dense clumps
\citep{wu05}:

\bq\label{wu05eq}
\rm SFR \sim 1.2 \times 10^{-8}
\left (\frac{M_{dense}}{M_{\sun}} \right) (M_{\sun} \ \rm yr^{-1}).
\eq

This result suggests that studying SFR--gas relations in our Galaxy
may be useful for interpreting star formation observed in nearby and
high-$z$ galaxies.  On the high--mass end of the spectrum, a large
survey of massive dense clumps by \citet{wu10}, provides a sample that
can be directly compared to extragalactic star formation tracers.

In this paper, we extend the comparison by \citet{evans09} by combining the 7
c2d clouds and 13 clouds from the GB survey. Regions
of high--mass star formation from a survey of $\sim$50 massive
dense Galactic clumps from \citet{wu10} provide an extension to high-mass
star formation regions.  The layout of this paper is
organized as follows.  We discuss low--mass star formation in the c2d
and GB clouds and describe how $\Sigma_{\rm gas}$ is derived from
extinction maps and estimate SFR surface densities ($\Sigma_{\rm
SFR}$) by YSO counts in Sections~\ref{sdent} and~\ref{sfysot},
respectively. In Section~\ref{sdenyso}, we separate clouds into evenly
spaced contour intervals of $\Sigma_{\rm gas}$, measuring the
$\Sigma_{\rm SFR}$ and $\Sigma_{\rm gas}$ in these intervals.
Section~\ref{diff} discusses the differences between Galactic and
extragalactic gas and SFR surface density relations.
$A_{V}$ and CO measurements of $\Sigma_{\rm gas}$ are compared in
Section~\ref{comap}.  We investigate whether massive star forming
regions behave differently from low--mass star forming regions in
Section~\ref{mass}.  The effects of averaging over whole galaxies
(kpc scales), including both star forming gas and diffuse
molecular gas, on the SFR--gas relation measured in extragalactic studies
are discussed in Section~\ref{res}.  Finally, we
summarize our results in Section~\ref{summary}.

\section{Low--mass Star Forming Regions from \spitzer \ c2d and Gould's Belt Surveys}\label{low}

The cores to disks (c2d) Legacy project included 5 large clouds:
Serpens (Ser), Perseus (Per), Chamaeleon II (Cha II), Ophiuchus (Oph),
and Lupus (Lup) \citep{evans09}. Because the Lup `cloud' is really
composed of several separate clouds, we divide them in this study by
name: Lup I, III, and IV, and obtain a total of 7 clouds.  The Gould's
Belt (GB) Legacy project (Allen \etal 2010, in preparation) includes
13 large clouds: IC 5146E and IC 5146NW \citep{harvey08}, Corona
Australis (CrA), Scorpius (Sco), Auriga (Aur), Auriga North (AurN),
Serpens-Aquila (Ser-Aqu), Musca (Mus), Cepheus (Cep) \citep{kirk09},
Cha I and III, and Lup V and VI.  These 20 clouds span a large range
of masses, areas, and number of YSOs (see Table~\ref{t1}). The term
``large" was used in the c2d study to distinguish them from the sample
of small clouds and cores that were biased toward regions known to
have dense gas \citep{evans03}. The``large" clouds are thus suitable
for statistical analyses, such as those presented here, but they are
actually small compared to the Orion cloud or many clouds in the inner
Galaxy.

\subsection{Estimating $\Sigma_{\rm gas}$ from Extinction Maps}\label{sdent}
We derive cloud masses ($M_{\rm gas, cloud}$) and mean surface
densities ($\Sigma_{\rm gas, cloud}$) from the extinction maps, which
were produced from a combination of 2MASS and \spitzer \ data, ranging
from 1.25$\micron$ to 24$\micron$.  In this wavelength range, the
spectral energy distributions (SEDs) of sources classified as stars
provide measurements of the visual extinction (Av) along lines of
sight through the clouds (Evans \etal 2007; Huard \etal 2010).
Line-of-sight extinctions were determined by fitting the SEDs,
adopting the \citet{weindraine01} extinction law with
$R_{V}$=$A_{V}/E(B-V)$=5.5. Extinction maps were constructed by
convolving these line--of--sight measures with uniformly spaced
Gaussian beams.  The c2d team observed ``off-cloud fields'' for four
of the molecular clouds: Chamaeleon, Perseus, Ophiuchus, and Lupus.
The line--of--sight extinction measurements from these off-cloud
fields suggested $A_{V}$ calibration offsets of 1--2 magnitudes;
therefore, in constructing the maps for these four clouds, Evans et
al. (2007) subtracted these calibration offsets.  Since no off--cloud
field had been observed for Serpens, they used a weighted mean of the
$A_{V}$ calibration offsets to correct the calibration of their
Serpens extinction maps.  No off-cloud fields were observed for the GB
survey. Further analysis of the fitting of the line-of-sight
extinctions demonstrates that the inferred calibration offsets
strongly depend on which wavebands had detections (Huard \etal 2010).
For example, sources with only near-infrared (2MASS) detections may
suggest no calibration offset, while sources with only mid-infrared
(\spitzer) \ detections show greater calibration offsets, perhaps as
high as 2--3 magnitudes.  This finding suggests that the
\citet{weindraine01} extinction law does not accurately characterize
the reddening through the full range 1.25--24 $\mu$m spectral
range. For if it did, the inferred $A_{V}$ calibration offsets should
be independent of the detected wavebands.  For this reason, the
extinction maps delivered by the GB survey make use of the catalogued
line--of--sight extinctions with no correction for potential
calibration offsets, and, for consistency, they suggest that the
previously adopted $A_{V}$ calibration ``offsets'' of 1--2 magnitudes
be added to the c2d extinction maps (Huard \etal 2010).  After
revising accordingly the extinctions in the clouds mapped by c2d, we
find that the gas masses, and thus the cloud surface densities, are
$\sim$20--30\% greater than those previously published by Evans et
al. (2009).  The extinction maps used in this study probe to higher
$A_{V}$ (up to 40 mag) than some previous studies (e.g.,
\citet{pineda08, lombardi08, lombardi10}) due to the inclusion of both
2MASS and mid--IR \spitzer \ data.

In order to compute the $M_{\rm gas, cloud}$, we chose extinction maps
with 270$\arcsec$ beams for all clouds.  We base this choice on the
best resolution map available for Ophiuchus, which is limited in
resolution due to a large extended region of high extinction with
relatively few background stars detected.  $M_{\rm gas, cloud}$ and
$\Sigma_{\rm gas, cloud}$ were calculated by summing up extinction map
measurements and converting to the column density using the relation
$N_{\rm H}/A_{V}$ = $(1.086C_{\rm ext}(V))^{-1}$ = 1.37$\times
10^{21}$ cm$^{-2}$ mag$^{-1}$ \citep{draine03} for a
\citet{weindraine01} $R_{V}$=5.5 extinction law, where $C_{\rm
ext}(V)$ = 6.715$\times 10^{-22}$ cm$^{2}$/H from the on-line
tables\footnote{Tables available at
http://www.astro.princeton.edu/$\sim$draine/dust/dustmix.html}, using
equation~\ref{mgas} and~\ref{siggas}, respectively. The uncertainties
in $M_{\rm gas, cloud}$ are computed from maps of extinction
uncertainty, which account for the statistical photometric
uncertainties, but not systematic uncertainties in using the
extinction law calibration.

We compute $M_{\rm gas, cloud}$ by summing up all pixels $\bigl (\sum
A_{V} \bigr)$ above $A_{V} =$ 2 in all clouds except for Serpens and
Ophiuchus which are covered by the c2d survey completely down to
$A_{V}$ = 6 and 3, respectively.  $M_{\rm gas, cloud}$ is then:

\begin{eqnarray}\label{mgas}
M_{\rm gas, cloud} = \mu \,m_{\rm H}\bigl (1.086C_{\rm ext}(V)\bigr
)^{-1} \times \sum
A_{V} \times {A_{\rm pixel}} \nonumber \\
\approx  1.58\times 10^{-36} \times \sum \ \biggl (\frac{A_{V}}{\rm
 mag}\biggr) \times \biggl (\frac{A_{\rm pixel}}{\rm
 cm^{2}}\biggr) \, (M_{\sun})
\end{eqnarray}

where the mean molecular weight ($\mu$) is 1.37, the total
number of hydrogen atoms is N(H)$\equiv$N(H{\sc i}) + 2N(H$_{2}$), and
we assume a standard molecular cloud composition of 63\% hydrogen,
36\% helium, and 1\% dust, $m_{\rm H}$ is the mass of hydrogen in
grams, the area of a pixel in square cm ($A_{\rm pixel}$) in the
extinction map is ($\pi$/180/3600)$^{2}$ D(cm)$^{2}$
R($\arcsec$)$^{2}$, where R($\arcsec$) is the pixel size in
arcseconds, and $A_{\rm cloud}$ is the area of the cloud measured in
square pc.  We divide $M_{\rm gas, cloud}$ by the area to obtain
$\Sigma_{\rm gas, cloud}$ for each cloud:

\begin{eqnarray}\label{siggas} 
\Sigma_{\rm gas,cloud}  = \left (\frac{M_{\rm gas, cloud}}{M_{\sun}}
\right ) \times \left (\frac{A_{\rm cloud}}{\rm pc^{2}}\right)^{-1}\
(M_{\sun} \ \rm pc^{-2}) \\
\Sigma_{\rm gas,cloud}  = 15 \biggl (\frac{A_{V}}{\rm mag}\biggr) (M_{\sun} \ \rm pc^{-2})  .
\end{eqnarray}

Measured cloud properties for c2d and GB clouds within a contour of
$A_{V} >$ 2 or $A_{V}$ completeness limit are shown in Table~\ref{t1}.

\subsection{Estimating Star Formation Rates from YSO Counts}\label{sfysot}

We estimate the SFR from the total number of YSOs ($N_{\rm YSO,tot}$)
contained in an area where $A_{V} > 2$, as described in
$\S$~\ref{sdent}.  We assume a mean YSO mass ($\langle M_{\rm YSO}
\rangle$) of 0.5$\pm$0.1 $M_{\sun}$, where the mean estimated error in
mass is derived from the mass distribution of YSOs in Cha II from
\citet{spezzi08}. The mean YSO mass is consistent with IMF studies by
\citet{chabrier03, kroupa02, nintra06}.  We also assume a period for
star formation ($t_{\rm Class \, II}$) of 2$\pm$1 Myr, based on the
estimate of the elapsed time between formation and the end of the
Class II phase \citep{evans09}.  This SFR assumes that star formation
has been continuous over a period greater than $t_{\rm Class \,
II}$. All clouds have Class III objects, indicating that star
formation has continued for longer than $t_{\rm Class \, II}$. The SFR
measured in this way could be underestimated or overestimated in any
particular cloud, but over an ensemble of 20 clouds, it should be the
most reliable SFR indicator available because no extrapolation from
the massive star tail of the IMF is needed.  We base our error
estimates by choosing the largest error from either the systematic
error, combined in quadrature from mean YSO mass and period of star
formation, or the Poisson error from YSO number counts.

\bq\label{ysosfr}
\begin{split}
\Sigma_{\rm SFR} = N_{\rm YSO, tot}\times \biggl (\frac{\langle M_{\rm
   YSO}\rangle}{M_{\sun}}\biggr ) \times \biggl (\frac{t_{\rm
   Class\,II}}{\rm Myr} \biggr)^{-1} \times \\
   \biggl (\frac{A_{\rm
cloud}}{\rm kpc^{2}} \biggr)^{-1} (M_{\sun} \, \rm yr^{-1} \, \rm kpc^{-2})
\end{split}
\eq

\begin{figure}
\epsscale{0.9}
\plotone{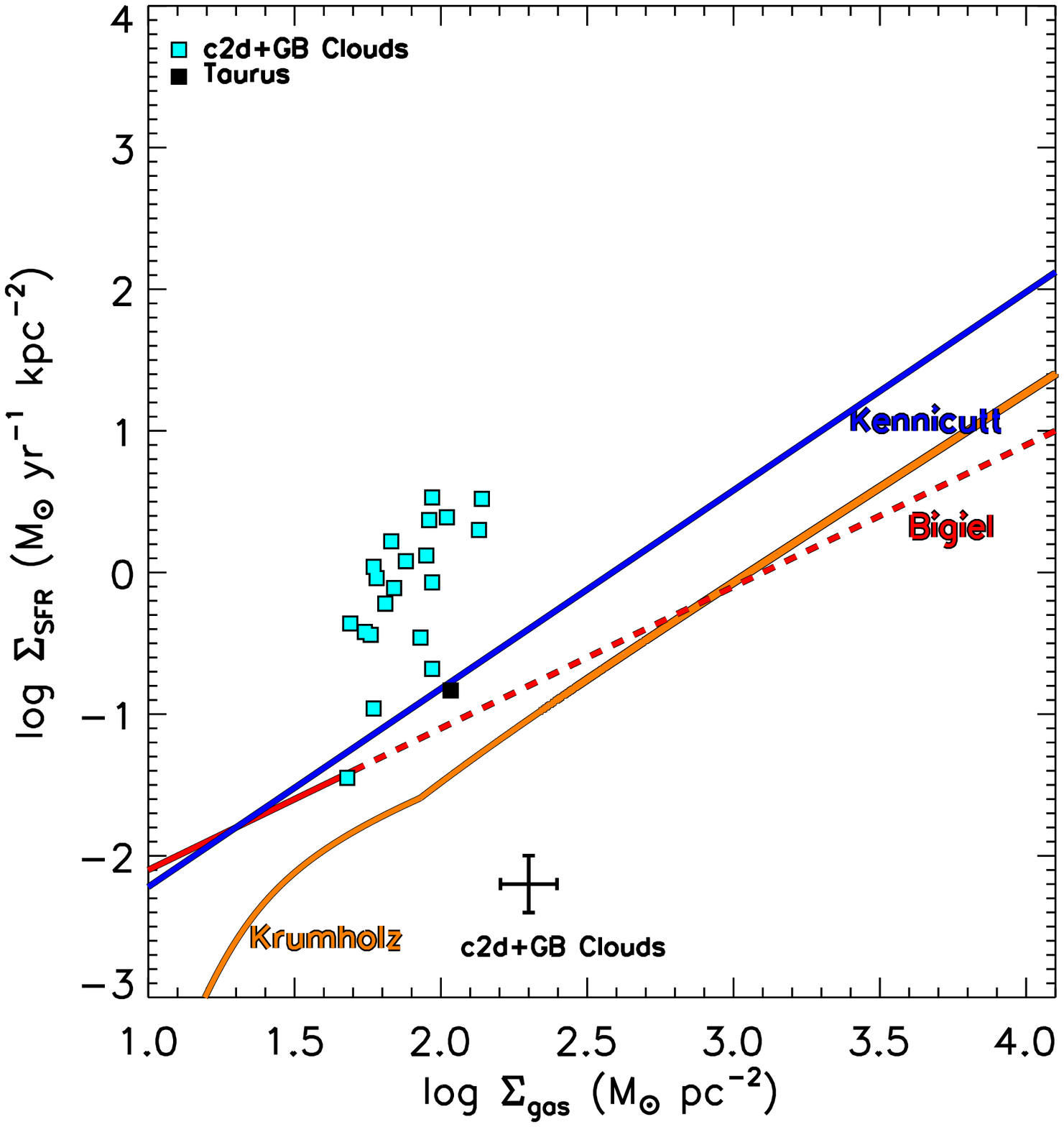}
\caption{$\Sigma_{\rm SFR}$ is shown versus the $\Sigma_{\rm
 gas}$ for c2d and GB clouds (cyan squares).  All cloud $\Sigma_{\rm
 gas}$ are measured above $A_{V}>2$ (or the cloud completeness limit,
 see Section~\ref{sdent}).
We also include an estimate for the Taurus molecular cloud (black
square) which includes YSO counts from \citet{rebull10} and an $A_{V}>2$ gas mass
from \citet{pineda10}.  Extragalactic observed
relations are shown for the sample of \citet{k98} and \citet{bigiel08}
(blue solid and red lines, respectively).  The
\citet{krumholz09} prediction for the total (\h1$+$CO) gas star
formation law for the galactic metallicity and a clumping factor of 1
corresponding to $\sim$100 pc scales is also shown (orange
line).
}
\label{ctot}
\end{figure}

Table~\ref{t1} lists values for clouds within a contour of $A_{V} >$~2
for all c2d and GB clouds. We show our estimated $\Sigma_{\rm gas,
cloud}$ and $\Sigma_{\rm SFR}$ for the c2d and GB clouds in
Figure~\ref{ctot}. $\Sigma_{\rm gas, cloud}$ ranges from $\sim$50--140
$M_{\sun} \ \rm pc^{-2}$, and $\Sigma_{\rm SFR}$ ranges from
$\sim$0.4-3.4 $M_{\sun} \ \rm kpc^{-2}\ \rm yr^{-1}$. We use these units
for convenience in comparing to the extragalactic relations.

We compare the observations to the predicted values for $\Sigma_{\rm
SFR}$ using $\Sigma_{\rm gas, cloud}$ that we calculate for the c2d
and GB clouds.  We plot these extragalactic relations in
Figure~\ref{ctot} and will include them in all the following SFR--gas
relation figures. The solid lines represent the regime where they were
fitted to data and the dashed lines are extrapolated relations
spanning the range of $\Sigma_{\rm gas}$. The blue line is from
disk-averaged or global SFR measurements based on H$\alpha$ emission
and the total (\h1$+$CO) gas surface densities in a sample of normal
spirals and starburst galaxies from \citet{k98}.  The red line is from
\citet{bigiel08}, who made sub-kpc resolution measurements in a sample
of spiral and dwarf galaxies using SFRs based on a combination of
\spitzer \ 24\micron \ and GALEX UV data and use CO measurements to
obtain a relation for H$_{2}$ gas surface density. Both of these studies
trace either obscured (24\micron) or unobscured (H$\alpha$ and UV)
massive star formation and are blind to regions of low--mass star
formation that we are measuring in this work.  We also compare to the
theoretical total (\h1 $+$ CO) gas and SFR relation of
\citet{krumholz09} (orange solid line).  This prediction takes into
account three factors: the conversion of atomic to molecular gas,
metallicity, and clumping of the gas.  For our comparisons, we choose
galactic solar metallicity and a clumping factor of 1, which
corresponds to clumping on 100 pc scales. We include data points for
the Taurus molecular cloud, including YSO counts from
\citet{rebull10}, $\Sigma_{\rm gas}$ from a 2MASS extinction map
\citep{pineda10}, and the total $^{13}$CO and $^{12}$CO gas mass from
\citet{goldsmith08}.

If we take the average $\Sigma_{\rm gas, cloud}$ defined as the total
$M_{\rm gas, cloud}$ divided by the total area ($A_{\rm cloud}$) in
pc$^{2}$, we would find that the average molecular cloud in this study
has a surface density of 91.5 $M_{\sun} \ \rm pc^{-2}$ and a
$\Sigma_{\rm SFR}$ of 1.2 $M_{\sun}\ \rm kpc^{-2}\ \rm
yr^{-1}$. Taking this average $\Sigma_{\rm gas, cloud}$ and
calculating what the extragalactic relations would predict for the
average cloud SFR surface density, we would get 0.13, 0.07, and 0.03
$M_{\sun} \ \rm kpc^{-2} \rm yr^{-1}$ for \citet{k98},
\citet{bigiel08}, and \cite{krumholz09}, respectively.  The observed
values exceed the observed extragalactic $\Sigma_{\rm SFR}$
predictions by factors of $\sim$9--17 and the theoretical prediction
by a factor of $\sim$40. While the star formation surface density,
$\Sigma_{\rm SFR}$ of 1.2 $M_{\sun}\ \rm kpc^{-2}\ \rm yr^{-1}$, seem
high, the clouds fill only a small fraction of the local square kpc.
From Table 1, the total SFR is 780.5 \msun \ Myr$^{-1}$. If we remove
the IC5146 clouds, which are more distant than 0.5 kpc, the SFR within
0.5 kpc is 748 \msun Myr$^{-1}$ or 7.5\ee{-4} \msunyr. Extrapolated to
the Galaxy with a star-forming radius of 10 kpc, this would amount to
0.3 \msunyr, less than the rate estimated for the entire galaxy of
0.68 to 1.45 \msunyr \citep{robitaille10}. This local, low-mass star
formation mode thus could account for a substantial, but not dominant,
amount of star formation in our Galaxy.

\subsection{Estimating $\Sigma_{\rm gas}$ and $\Sigma_{\rm SFR}$ for the Youngest YSOs as a Function of $A_{V}$}\label{sdenyso}

The last section gave us estimates over the whole molecular cloud
including all YSOs in each cloud.  Early work surveying large areas of
clouds (e.g., \citet{lada92}) suggested that star formation is
concentrated in regions within molecular clouds in regions of high
densities ($n\sim10^{4}$ \cmv).  The c2d and GB studies of many whole
clouds have clearly established that star formation is not spread
uniformly over clouds, but is concentrated in regions at high
extinction. Furthermore, the youngest YSOs and dense cores
\citep{enoch07} are the most highly concentrated at high $A_{V}$
(Evans \etal 2009, Bressert \etal 2010, submitted).  Older YSOs can
leave their original formation region or even disperse the gas and
dust. Taking the average velocity dispersion of a core to be 1 km
s$^{-1}$, a 2 Myr old YSO could travel $\sim$2 pc, roughly the average
radius of a cloud in this study.  We therefore apply a conservative
approach and only estimate the SFRs using the youngest Class I or Flat
SED YSOs (see \citet{greene94} for the definition of classes) that
have not yet migrated from their birth place. To classify YSOs as
Class I or Flat SED, we use the extinction corrected spectral index
from \citet{evans09} for the c2d clouds and the uncorrected spectral
index for the GB clouds (Table~\ref{t1}).  These two classes of YSOs
have timescales of 0.55$\pm$ 0.28 and 0.36$\pm$ 0.18 Myr, respectively
(Allen \etal 2010, in preparation).

\begin{figure}
\epsscale{1.1}
\plotone{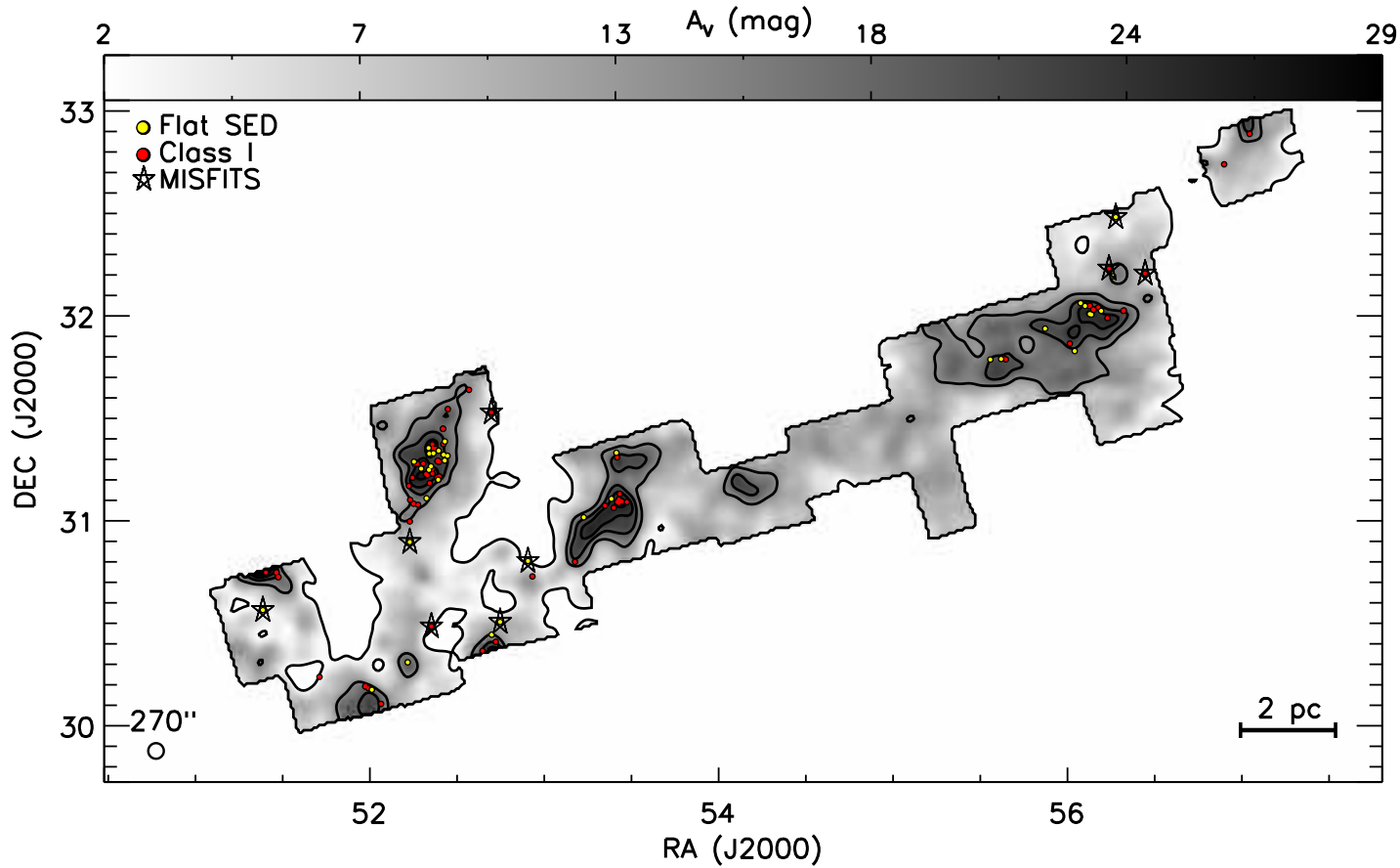}
\caption{ An example of the $\Sigma_{\rm gas}$ measurement method in
 the Perseus molecular cloud from the c2d survey.  The grayscale
 image is the extinction map with black contours ranging from 2--29
 in intervals of 4.5 mag The yellow filled circles are Flat SED
 sources and the red filled circles are Class I sources.  Sources
 that have an open star correspond to suspicious YSOs (MISFITS) that
 were observed in \hcop$J$=3--2 \ at the CSO and were not detected.
 We measure the $\Sigma_{\rm gas}$ from each map in each contour of
 extinction.  Contours are spaced in intervals wider than the
 extinction map beam size of 270$\arcsec$.  To estimate SFR, we
 count the YSOs in corresponding contour levels (Section~\ref{sfysot}).
 }
\label{cont}
\end{figure}

In order to measure $\Sigma_{\rm SFR}$ and $\Sigma_{\rm gas}$ for the
youngest YSOs, we divide the clouds into equally spaced contour levels
of $A_{V}$ or $\Sigma_{\rm gas, con}$ and measure the SFR, mass
($M_{\rm con}$), and area ($A_{\rm con}$) enclosed in that contour
level.  The contour intervals start from the extinction map
completeness limits ($\S$~\ref{sdent}) and are spaced such that they
are wider than our map beam size of 270$\arcsec$ as shown in
Figure~\ref{cont}.  We compute the gas surface density ($\Sigma_{\rm
gas, con}$) in the same way as in equation~\ref{siggas}, but this time using
only the mass ($M_{\rm gas, con}$) and area ($A_{\rm con}$) enclosed
in the $A_{V}$ contour region.

\bq\label{eq5} \Sigma_{\rm gas,con}  = \left (\frac{M_{\rm gas,
   con}}{M_{\sun}} \right ) \times \left (\frac{A_{\rm
con}}{\rm pc^{2}}\right)^{-1}\ (M_{\sun} \ \rm pc^{-2}).
\eq

If there are no YSOs found in the contour region, we compute an upper
limit to the SFR by assuming that there is one YSO in that region.
The upper limits are denoted by the asterisks in Table~\ref{tcont}.
We estimate the uncertainties in the both the SFR and $\Sigma_{\rm
SFR}$ by choosing the largest error: either the systematic or Poisson
error from YSO counts.

\subsubsection{MISidentified YSOs from SED FITS (MISFITS)}\label{misfit}

The c2d and GB surveys have classified YSOs based on the SED slope
from a fit to photometry between 2$\micron$ and 24$\micron$ (Evans
\etal 2009; Allen \etal 2010, in preparation). However, we find that
some of the Class I and Flat SED YSOs are not clustered and lie
farther from the extinction peaks than expected for their age.  Most
of these suspicious objects are found, on average, to lie at
$A_{V}\sim$6 magnitude.  If these Class I and Flat SED objects are
true young YSOs, they are more likely to be centrally concentrated toward
the densest regions in a cloud \citep{lada92}.  Class I and Flat SED
YSOs should be associated with a dense, centrally concentrated,
envelope of gas. We therefore performed a follow up survey of these
suspicious objects for a subset of c2d and GB clouds using the Caltech
Submillimeter Observatory (CSO). Our work was motivated by a study
performed by \citet{vankempen09}, who mapped \hcop$J$=4--3 using the
James Clerk Maxwell Telescope and found that 6 previously classified
Class I YSOs in Ophiuchus had no detections down to 0.1 K.

With high effective and critical densities, n$\sim$10$^{4}$ and
$\sim$10$^{6}$ cm$^{-3}$ \citep{evans99}, the \hcop$J$=3--2 \
transition provides a good tracer of the dense gas contained in
protostellar envelopes. We observed Class I and Flat SED YSOs at the
CSO from the Aur, Cep, IC5146, Lup, Oph, Per, Sco, Ser, and Ser--Aqu
molecular clouds using the \hcop$J$=3--2 (267.557620 GHz) line
transition as a dense stage I gas tracer to test if they are truly
embedded YSOs.  A survey of all observable c2d and GB clouds with
detailed results will be published in a later paper.

Observations were made during June and December of 2009 and July 2010
with an atmospheric optical depth ($\tau_{225}$) ranging from
0.05--0.2.  We observed each source using position switching for an
average of 120 seconds on and off source.  If a source was detected,
we integrated until we reach a signal--to--noise of $\geq 2\sigma$
(most sources have $\geq 3\sigma$ detections). Using an average main
beam efficiency ($\eta$) of 0.80 and 0.61 for December 2009 and July
2010 observing runs, respectively, we compute the main beam
temperatures and integrated intensities of detected sources. The
results are shown in Table~\ref{tmis}.

For this paper, we observed a total of 98 suspicious sources, 45 Flat
SED and 53 Class I sources.  We find that 74\% (73/98) of the observed
sources are not detected in \hcop$J$=3--2.  Out of the 42 Flat SED
sources, we detect only 3, but we detect 42\% (22/53) of the Class I
sources. The YSO MISFITS are a small fraction of the total number
(3146) of YSOs or Class I plus Flat sources (681) in the c2d and GB
studies, but they could bias the statistics upward at low gas surface
densities. The undetected MISFITS may be background galaxies or later
stage YSOs, and we will explore this in more detail in a later
paper. Figure~\ref{cont} shows the distribution of Class I (red filled
circles), Flat SED YSOs (yellow filled circles) and non--detected
MISFITS, indicated by the open stars on the Per cloud $A_{V}$ map.
MISFITS that we do not detect in \hcop$J$=3--2 are removed from the
sample when we measure $\Sigma_{\rm gas, con}$ and $\Sigma_{\rm SFR}$
in $A_{V}$ contours (Section~\ref{ysocont}).

\subsection{Results: The Youngest YSOs as a Function of $\Sigma_{\rm gas}$}\label{ysocont}

\begin{figure}
\epsscale{0.9}
\plotone{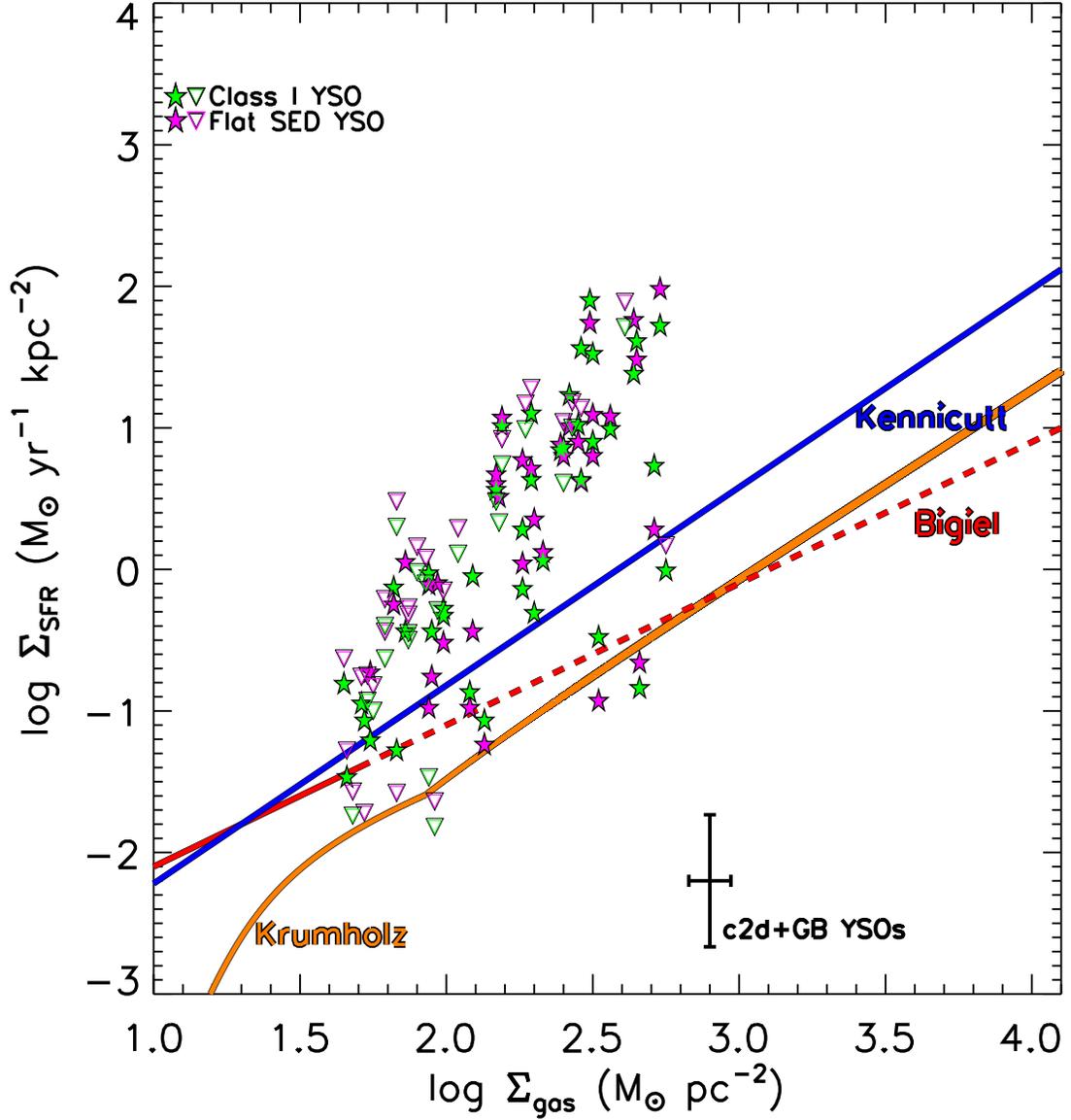}
\caption{Gas surface densities measured from extinction maps and SFRs
 estimated from Class I (green stars) and Flat SED (magenta stars)
 YSO number counts in c2d and Gould's belt clouds are shown.  For
 contour levels that do not contain any YSOs, we calculate an upper
 limit for that region using one YSO (open inverted
 triangles). Extragalactic observed relations are shown for the
 sample of \citet{k98} and \citet{bigiel08} (blue solid and red
 lines, respectively).
}
\label{yso}
\end{figure}

After removing the MISFITS from our Class I and Flat SED YSO sample,
we show the number, $M_{\rm gas, con}$, $\Sigma_{\rm gas, con}$, SFRs,
and $\Sigma_{\rm SFR}$ in Table~\ref{tcont} for all contour levels in
each cloud or separate cloud component (see Section~\ref{low}). In
Figure~\ref{yso}, we show the $\Sigma_{\rm gas, con}$ and $\Sigma_{\rm
SFR}$ densities for both Class I and Flat SED sources (green and
magenta stars) and upper limits for each class (green and magenta
inverted triangles) that we measured in contour regions described in
$\S$~\ref{sdenyso}, with extragalactic observational relations
over--plotted.  A wider range in both $\Sigma_{\rm gas}$
($\sim$45--560 $M_{\sun}\ \rm pc^{-2}$), and $\Sigma_{\rm SFR}$
($\sim$0.03--95 $M_{\sun}\ \rm kpc^{-2}\ \rm yr^{-1}$) are found for
contour regions compared to the total cloud measurements.  We note
that the points for Sco and Cep \citep{kirk09} clouds are obtained by
co-adding the separate cloud regions using the same contour
intervals. Since these points sample regions with non--uniform
$A_{V}$, they only provide an estimate of $\Sigma_{\rm SFR}$ and
$\Sigma_{\rm gas}$.  These points lie at $\Sigma_{\rm SFR} <$ 6
$M_{\sun}\ \rm kpc^{-2}\ \rm yr^{-1}$ and high $\Sigma_{\rm gas}$
$>$330 $M_{\sun}\ \rm pc^{-2}$.

We compare our YSO contour results to extragalactic relations and find
that most points lie well above the extragalactic relations.
Excluding upper limits, the mean values of $\Sigma_{\rm SFR}$ and
$\Sigma_{\rm gas, con}$ are of 9.7 $M_{\sun}\ \rm kpc^{-2}\ \rm
yr^{-1}$ and 225 $M_{\sun} \rm pc^{-2}$, respectively.  Evaluated at
this mean gas surface density, the extragalactic relations
under--predict $\Sigma_{\rm SFR}$ by factors of $\sim$21--54. The mean
YSO contour lies above the \citet{krumholz09} extragalactic SFR--gas
relation prediction by $\sim$2 orders of magnitude.  We explore the
differences between the Galactic and extragalactic SFR--gas relations
in Section~\ref{diff}.

\section{Why are Galactic SFR--gas Relations different from Extragalactic Relations?}\label{diff}

The differences between our findings on Galactic scales and the
 extragalactic relations, both on global or disk--averaged scales
 \citep{k98} and scales of hundreds of pc \citep{kennicutt07,
 thilker07,bigiel08,blanc09, braun09, verley10}, might be explained in
 the following ways. Firstly, using $^{12}$CO to measure the H$_{2}$
 in galaxies might give systematically different $\Sigma_{\rm gas}$
 than do $A_{V}$ measurements (Section~\ref{comap}). Secondly, the
 local c2d and GB clouds are forming low--mass stars; since
 extragalactic SFR tracers respond only to massive stars, the two star
 forming regimes might behave differently. In Section~\ref{mass}, we
 will investigate whether massive star forming regions agree with the
 extragalactic SFR--gas relations and if they vary from low--mass star
 forming regions. Finally, averaging over whole galaxies on scales
 of hundreds of pc includes both gas contained in the parts of
 molecular clouds that are forming stars and diffuse molecular gas
 that is not forming stars (Section~\ref{res}). A local example of
 this is a study of the Taurus molecular cloud; \citet{goldsmith08},
 found a large amount of diffuse $^{12}$CO at lower gas densities
 where no young stars are forming.  Extragalactic studies averaging
 over hundreds of pc scales would include this gas, causing an
 increase in the amount of CO flux that is being counted as star
 forming gas.

\subsection{The use of CO versus $A_{V}$ to determine $\Sigma_{\rm gas}$}\label{comap}

Since extinction maps are direct probes of $\Sigma_{\rm gas}$, they
provide the best measure of the total gas and are optimal for use in
determining the $\Sigma_{\rm gas}$ of molecular clouds. However,
$A_{V}$ maps are not easily obtainable in extragalactic studies, which
instead employ CO maps, particularly $^{12}$CO $J$=1--0, to determine
$\Sigma_{\rm gas}$ of molecular hydrogen.  Since the molecular
hydrogen (H$_{2}$) rotational transitions require high temperatures
not found in the bulk of molecular clouds, other tracers of dense gas
are used to estimate the amount of H$_{2}$.  The next most abundant
molecule with easily observable excitation properties in a molecular
cloud is $^{12}$CO $J$=1--0. In this study, we want to explore how well
CO traces $A_{V}$ as a function of $M_{\rm gas}$ or $\Sigma_{\rm
gas}$.  We can directly test this in two galactic clouds, Perseus and
Ophiuchus, which both have $^{12}$CO $J$=1--0 and $^{13}$CO $J$=1--0
maps from the Five College Radio Astronomy Observatory (FCRAO)
COordinated Molecular Probe Line Extinction Thermal Emission
(COMPLETE) Survey of Star Forming Regions \citep{ridge06}.

In order to directly compare the CO maps from the COMPLETE survey to
the $A_{V}$ maps in this study, we interpolate the CO data onto the
$A_{V}$ map grid with a pixel size of 45$\arcsec$.  We integrate the
publicly available CO data cubes over the velocity range from 0--15 km
s$^{-1}$ to create moment zero maps of integrated intensity defined
as: $I_{\rm CO} \rm (x,y) \equiv \int T_{\rm mb}\rm (x,y,z) dV \rm \ K
\rm \ km \ \rm s^{-1}$, where $T_{\rm mb}\equiv T_{\rm
A}^{*}/\eta_{\rm mb}$ is the main beam brightness temperature defined
as the antenna temperature ($T_{\rm A}^{*}$) divided by the main beam
efficiency ($\eta_{\rm mb}$) of 0.45 and 0.49 for $^{12}$CO $J$=1--0
and $^{13}$CO $J$=1--0, respectively, from \citet{pineda08}.
Corresponding rms noise maps ($\sigma_{\rm T(x,y)}$) were constructed
by calculating the standard deviation of intensity values within each
spectroscopic channel where no signal is detected. In order to
determine the gas surface density of H$_{2}$ ($\Sigma_{\rm H_{2}}$),
we must first compute the column density of H$_{2}$.  The column
density of H$_{2}$ is estimated from the $^{12}$CO map by using a
CO--to--H$_{2}$ conversion factor ($X_{\rm CO}$), which is defined as
the ratio of H$_{2}$ column density to the integrated intensity
($X_{\rm CO} \equiv N_{\rm H_{2}}/I_{\rm CO}$). Similarly for the
$^{13}$CO map, the column density is derived from the $^{12}$CO and
$^{13}$CO maps, assuming LTE and an abundance ratio of
H$_{2}$--to--$^{13}$CO.  To compare $\Sigma_{\rm H_{2}}$ from
$^{12}$CO and $^{13}$CO to the $\Sigma_{\rm gas}$ from $A_{V}$, we use
only regions in the CO maps that have emission lines with positive
integrated intensities and line peaks that are greater than 5 times
the rms noise.  Our masses from extinction measurements used for this
comparison are therefore slightly lower than those in Tables~\ref{t1}
and~\ref{tcont} by $\sim$5\% (See Tables~\ref{tco} and~\ref{tco2}).

The $X_{\rm CO}$ factor for $^{12}$CO has been derived using a variety
of methods such as gamma ray emission caused by the collision of cosmic rays
with hydrogen \citep{bloemen86}, virial mass methods \citep{solomon87,
blitz07}, maps of dust emission from $IRAS$ and assuming a constant
dust--to--gas ratio \citep{frerking82}, extinction maps from optical
star counts \citep{duvert86,bachiller86,langer89} and 2MASS data
\citep{lombardi06, pineda08}, and theoretically by the assumption that
giant molecular clouds are in gravitational equilibrium
\citep{dickman86,heyer01}.  All these studies find a range of $X_{\rm
CO}$ of 0.9--4.8$\times 10^{20}$ cm$^{-2}$ K$^{-1}$ km$^{-1}$ s, but
they were almost all restricted to regions with $A_{V} < 6$ mag.
Studies of extragalactic SFR--gas relations chose values close to the
average galactic $X_{\rm CO}$ measurements in the literature:
2.0$\times 10^{20}$ cm$^{-2}$ K$^{-1}$ km$^{-1}$ \citep{bigiel08}) or
2.8$\times 10^{20}$ cm$^{-2}$ K$^{-1}$ km$^{-1}$ s from
\citet{bloemen86} \citep{k98, kennicutt07,blanc09}.  Since the goal of
this study is to compare to extragalactic measurements, we choose a
$X_{\rm CO}$ of 2.8$\pm$0.7$\times 10^{20}$ cm$^{-2}$ K$^{-1}$
km$^{-1}$ s from \citet{bloemen86} to be consistent with the study of
\citet{k98}.

We compute the column density of H$_{2}$ from $^{12}$CO measurements
using the equation:

\bq
N_{\rm H_{2}(^{12}\rm CO)} = X_{^{12}\rm CO} \times I_{^{12} \rm CO} \ (\rm cm^{-2}).
\eq

This can be rewritten in terms of gas surface density:

\begin{eqnarray}
\begin{split}
\Sigma_{\rm H_{2}(^{12}\rm CO)} = \bigl (2 m_{\rm H}\times N_{\rm
 H_{2}(^{12}\rm CO)} \times A_{\rm pixel} \bigr) / A_{\rm cloud}\nonumber \\
\approx  10^{-33}\times m_{\rm H} \times \biggl(\frac{N_{\rm
 H_{2}(^{12}\rm CO)}}{\rm cm^{-2}}\biggr)\times \biggl (\frac{A_{\rm pixel}}{\rm
 cm^{2}}\biggr) \times \\
 \biggl (\frac{A_{\rm cloud}}{\rm pc^{2}}\biggr)^{-1}  \ (M_{\sun} \ \rm pc^{-2}),
\end{split}
\end{eqnarray}

where we take the total number of hydrogen atoms is
N(H)$\equiv$2N(H$_{2}$).  We use this factor of two instead of the
mean molecular weight of H$_{2}$ ($\mu_{\rm H_{2}}$=2.8, derived from
cosmic abundances of 71\% hydrogen, 27\% helium, and 2\% metals, e.g.,
\citet{kauffmann08}) to be consistent with the extragalactic studies
of \citet{k98} and \citet{bigiel08}.  This factor of 2 does not
account for helium, which is an additional factor of $\sim$1.36
\citep{hildebrand83}.  The errors in our gas surface density
measurements include both the error from the rms intensity maps and the error in the CO--to--H$_{2}$ conversion factor from
\citet{bloemen86}.

\begin{figure}
\epsscale{0.95}
\plotone{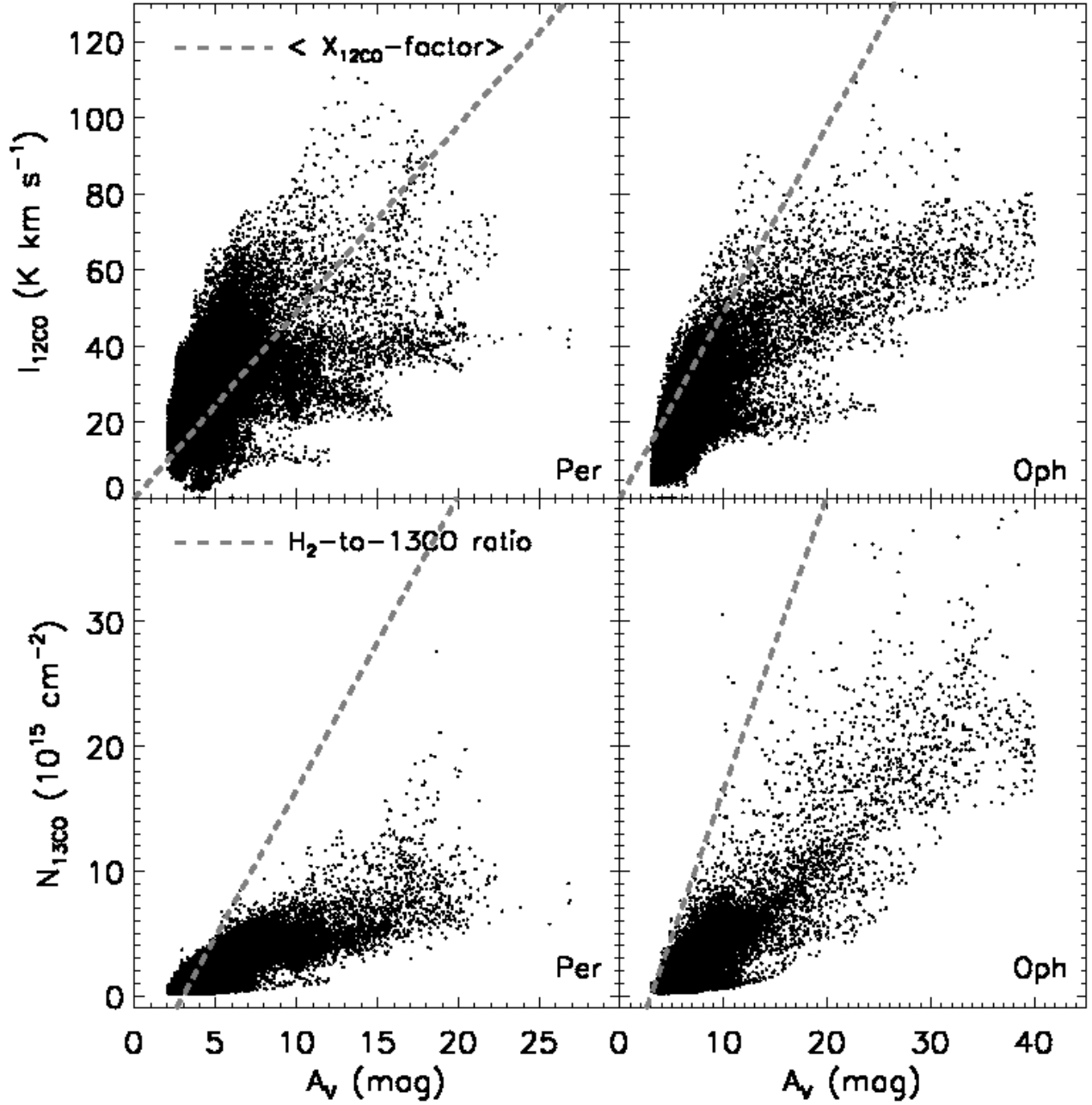}
\caption{{\bf top panels:} $^{12}$CO integrated intensity versus
 visual extinction ($A_{V}$) for Per (left) and Oph (right). The
 standard $X_{\rm CO}$--factor fit from \citet{bloemen86} is shown by
 the dashed grey lines (Section~\ref{comap}).  {\bf bottom panels:}
 $^{13}$CO column densities versus visual extinction ($A_{V}$) for
 Per (left) and Oph (right). The average H$_{2}$--to--$^{13}$CO
 abundance ratio from the literature is shown by the grey dashed
 lines (Section~\ref{comap}).  }
\label{fcofit1}
\end{figure}

The top two panels of Figure~\ref{fcofit1} show the $^{12}$CO
integrated intensity versus the visual extinction derived from the
2MASS and \spitzer \ data for both Per (left) and Oph (right). We
over--plot the conversion factor derived from the gamma ray study of
\citet{bloemen86} (dashed line).  $^{12}$CO is seen to correlate with
$A_{V}$ out to $A_{V}\sim$7--10, where $^{12}$CO starts to saturate
and the distribution flattens out to higher $A_{V}$.  A large
difference in the amount of $^{12}$CO integrated intensity produced
relative to that predicted by $X_{^{12}\rm CO}$ between the Per and Oph
molecular clouds is seen.  This may be due to higher opacity
at higher $A_{V}$ in Oph relative to Per. Extinction
values around 10--20 mag are found to be essentially invisible to
$^{12}$CO.  This figure demonstrates the non--linear, non--monotonic
behavior of CO emission with $A_{V}$.

$^{12}$CO, however, is not the most reliable tracer of star forming
gas because of high opacity and varying
$^{12}$CO--to--H$_{2}$ abundance due to photodissociation or depletion
on to dust grains.  Studies of molecular clouds \citep{carpenter95,
heyer96,goldsmith08} show that $^{12}$CO contains a significant
diffuse component in the low column density regime $A_{V}
<$4. $^{13}$CO emission is a more reliable tracer of dense gas ranging from
1000-7000 cm$^{-3}$ than $^{12}$CO because it is optically thin for
most conditions within a cloud and $^{13}$CO abundance variations are
small for densities $<$5000 cm$^{-3}$ and temperatures of $>$15 K
\citep{bachiller86,duvert86,heyer95,caselli99}.

To estimate the column densities from the $^{13}$CO $J$=1--0
integrated intensity maps we assume local thermodynamic equilibrium
(LTE), optically thin $^{13}$CO $J$=1--0, and that $^{12}$CO $J$=1--0
and $^{13}$CO $J$=1--0 have equivalent excitation temperatures.  In
order to derive column densities, we also determine an optical depth
($\tau_{^{13}\rm CO}$) and excitation temperature ($T_{\rm ex}$) from
comparison to the $^{12}$CO $J$=1--0 line.  We can derive this by
assuming the $^{12}$CO $J$=1--0 is optically thick; as $\tau\to\infty$,

\bq
T_{\rm ex} = \frac{5.5}{{\rm ln}\left (1 +  \frac{5.5}{T_{peak,^{12}\rm
     CO} + 0.82} \right )} \ (\rm K),
\eq

where $T_{\rm peak,^{12}\rm CO}$ is the $^{12}$CO $J$=1--0 peak main beam brightness temperature which we measure on a pixel--by--pixel basis.

The $^{13}$CO $J$=1--0 optical depth is then

\bq
\tau_{^{13}\rm CO} = -{\rm ln} \left(1 - \frac{T_{\rm peak,^{13}{\rm
   CO}}}{5.3}\left [\frac{1}{\exp^{(5.3/T_{\rm ex})} -1} -0.16
 \right ]^{-1}\right ),
\eq

where $T_{\rm peak,^{13}\rm CO}$ is the $^{13}$CO $J$=1--0 peak main
beam brightness temperature measured in each pixel.  We can then use the definition of $^{13}$CO
optical depth and column density from \citet{rw96} to estimate the
column density of $^{13}$CO:

\bq
\begin{split}
N_{^{13}\rm CO} = 2.6\times 10^{14} \left ( \frac{\tau_{^{13}\rm CO}}{1
- \exp^{-\tau_{^{13}\rm CO}}} \right ) \times \\ 
\left (\frac{ I_{^{13}\rm CO}}{1 - \exp^{(-5.3/T_{\rm
     ex})}}\right ) \ (\rm cm^{-2}).
\end{split}
\eq

Certain regions near $A_{V}$ peaks in the clouds are optically thick
and are affected by $^{12}$CO self absorption.  In these regions, we
cannot accurately determine $T_{\rm ex}$ and therefore $\tau_{^{13}\rm
CO}$. This problem affects $\sim$10\% of the pixels in Oph ($A_{V} >$
30 mag) and $\sim$5\% of the pixels in Per ($A_{V} >$ 20 mag).  Since
most of the mass lies at low $A_{V}$, we mask out the pixels that have
$^{13}$CO self absorption and do not include them in determining
the $^{13}$CO column density and mass estimate discussed below.

For comparison with extragalactic work, where clouds are not resolved,
we derive $^{13}$CO column densities by using average spectra to
determine $T_{\rm ex}$ and $\tau_{^{13}\rm CO}$ using a $T_{\rm
peak,^{12}\rm CO}$ of 3 and 5.3 K and a $T_{\rm peak,^{13}\rm CO}$ of
0.7 and 1.5 K for Per and Oph, respectively.
Comparing the two methods, we find that using peak
temperatures from average spectra or a constant $T_{\rm ex}$ and
$\tau_{^{13}\rm CO}$ will result in higher $N_{^{13}\rm CO}$ by a
factor of $\sim$2 at $A_{V}<$ 10 over the pixel--by--pixel
measurements.

In order to convert the $N_{^{13}\rm CO}$ column density into H$_{2}$
column density, we use a H$_{2}$--to--$^{13}$CO abundance ratio. The
H$_{2}/^{13}$CO abundance ratio for the Per cloud was determined by
\citet{pineda08}, who found a value of 3.98$\pm$0.07$\times 10^{5}$
for $A_{V} < 5$ mag.  Dividing the cloud into separate regions, they
found an average abundance ratio of 3.8$\times 10^{5}$. Other values
found in the literature range from 3.5--6.7$\times 10^{5}$, with an
average value of $\sim$4$\times 10^{5}$ using both extinction maps
\citep{pineda08} and star counts
\citep{bachiller86,duvert86,langer89}.  We adopt the average
$H_{2}/^{13}$CO ratio from the literature of (4$\pm$0.4)$\times
10^{5}$ to convert $^{13}$CO to H$_{2}$ column densities using the
relation:

\bq
N_{\rm H_{2}(^{13}\rm CO)} =  (4\pm0.4)\times 10^{5} N_{^{13} \rm CO} \ (\rm cm^{-2})
\eq

or in terms of surface densities:

\begin{eqnarray}
\begin{split}
\Sigma_{\rm H_{2}(^{13}\rm CO)} = \bigl (2 m_{\rm H}\times N_{\rm
 H_{2}(^{13}\rm CO)} \times A_{\rm pixel} \bigr) / A_{\rm cloud}\nonumber \\
\approx  10^{-33}\times m_{\rm H} \times \biggl(\frac{N_{\rm
 H_{2}(^{13}\rm CO)}}{\rm cm^{-2}}\biggr)\times \biggl (\frac{A_{\rm pixel}}{\rm
 cm^{2}}\biggr) \times \\
 \rm \biggl (\frac{A_{\rm cloud}}{\rm pc^{2}}\biggr)^{-1}  \ (M_{\sun} \ \rm pc^{-2}),
\end{split}
\end{eqnarray}

where we choose a factor of two instead of the mean molecular weight
in order to consistently compare to $^{12}$CO.  We show the $^{13}$CO
integrated intensity versus the visual extinction derived from the
2MASS and \spitzer \ data for both Per (left) and Oph (right) in the
bottom two panels of Figure~\ref{fcofit1}.  The average
$H_{2}/^{13}$CO ratio is shown by the dashed line. A turnover in
$^{13}$CO is seen at $A_{V}\sim$7 (Per) and $\sim$10 (Oph) that is
likely due to an increase in optical depth.  The amount of $^{13}$CO
in Per follows the average abundance ratio out to $A_{V}\sim$5, but in
Oph, the $^{13}$CO integrated intensity is underproduced.

\begin{figure}
\epsscale{0.9}
\plotone{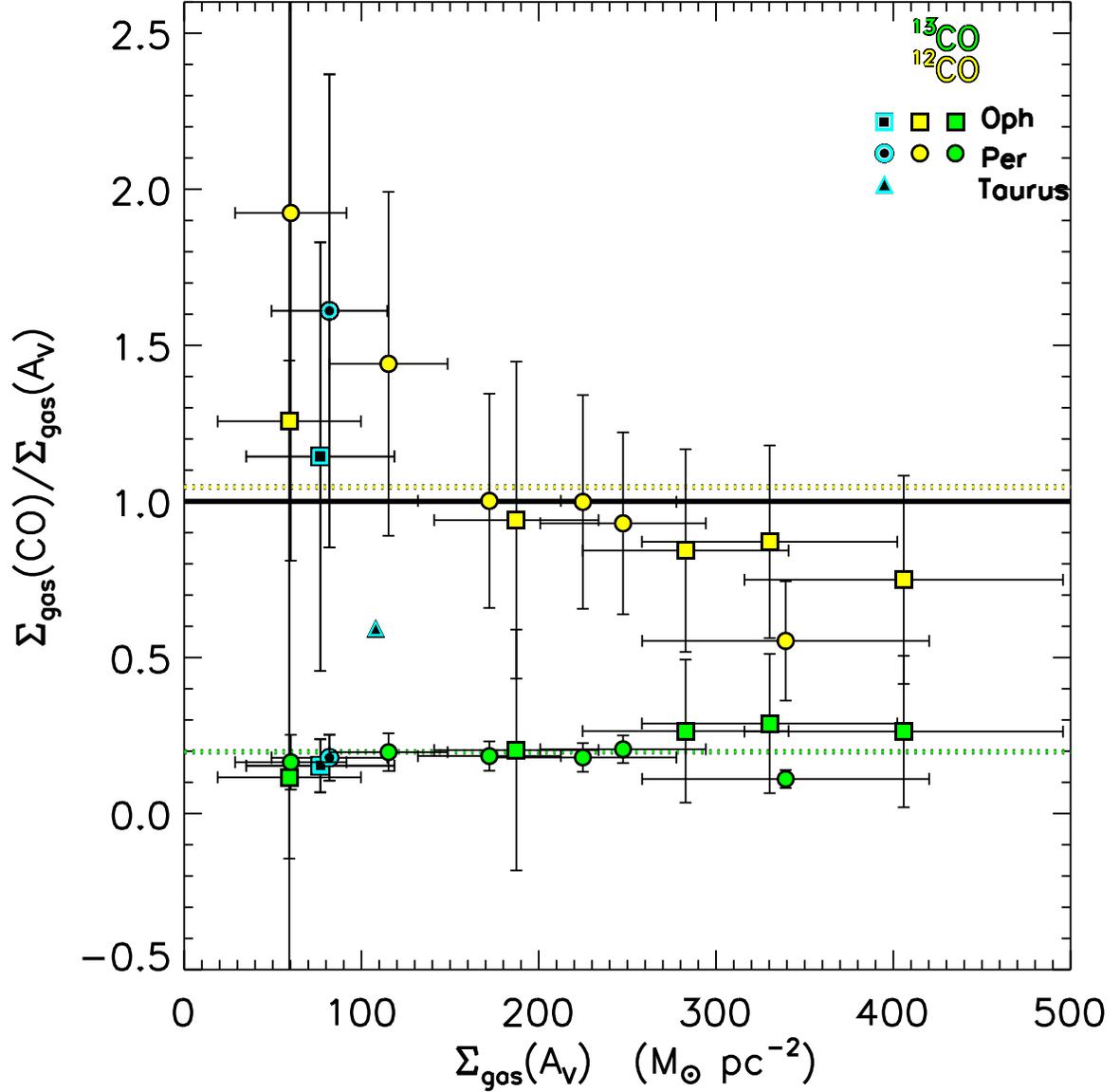}
\caption{The ratio of H$_{2}$ gas surface densities from CO compare to
 that estimated from $A_{V}$ maps ($\Sigma_{\rm gas}$). The cyan
 squares and circles are points for the Oph and Per clouds,
 respectively. The filled green ($^{13}$CO ) and yellow ($^{12}$CO)
 squares (Oph) and circles (Per) are measurements in evenly spaced
 contour intervals of $A_{V}$. The dashed horizontal green and yellow
 lines are the average of $^{13}$CO and $^{12}$CO contour points.  If
 CO traces the mass we find using extinction maps, we would expect the
 ratio of CO/$A_{V}$ to be of order unity as shown by the solid black
 line.}
\label{fco2}
\end{figure}

To test how well CO traces $A_{V}$ as a function of $M_{\rm gas}$ or
$\Sigma_{\rm gas}$, we measure $\Sigma_{\rm gas}$ densities in our
$A_{V}$ maps and $\Sigma_{\rm H_{2}}$ in the CO maps in the
overlapping area where there is a positive CO integrated intensity
over 5 times the rms noise.  In Figure~\ref{fco2}, we plot the ratio
of $\Sigma_{\rm H_{2}}$ and $\Sigma_{\rm gas}$ from $A_{V}$, which are
effectively mass ratios, since the area measured is the same.  The
cyan squares and circles are points for the c2d and GB clouds
($\Sigma_{\rm H_{2}(^{12}\rm CO, cloud)} , \Sigma_{\rm H_{2}(^{13}\rm
CO, cloud)}$) and the filled green and yellow squares and circles are
measurements in contours of $A_{V}$ using the same method as in
Section~\ref{sdenyso} ($\Sigma_{\rm H_{2}(^{12}\rm CO, con)} ,
\Sigma_{\rm H_{2}(^{13}\rm CO, con)}$) for Oph and Per, respectively
(Tables~\ref{tco} and ~\ref{tco2}). A measurement for the Taurus cloud
using both $^{12}$CO and $^{13}$CO above $A_{V}=2$ from
\citep{goldsmith08} is also shown (cyan triangle).  If CO traces the
mass we find using extinction maps, we would expect the ratio of
CO/$A_{V}$ mass to be of order unity as shown by the solid black line
in Figure~\ref{fco2}.  For $^{12}$CO, we find the total cloud
measurement for Per to have $\Sigma_{\rm H_{2}}$ of $\sim
1.6\Sigma_{\rm gas}$ at $\Sigma_{\rm gas} \ltsim$ 100 $M_{\sun} \ \rm
pc^{-2}$, but the ratio is close to unity within the errors.  We find that
$^{12}$CO traces $A_{V}$ relatively well in the Oph cloud out to
$\sim$200 $M_{\sun} \ \rm pc^{-2}$.  At $\Sigma_{\rm gas} \gtrsim 200
M_{\sun} \ \rm pc^{-2}$, $^{12}$CO underestimates the $A_{V}$ mass in
both Per and Oph by $\sim$ 30\%, on average.

Since $^{13}$CO should trace denser gas \citep{duvert86,bachiller86},
we also explore how it traces $A_{V}$ as a function of $\Sigma_{\rm
gas}$.  We plot this on Figure~\ref{fco2} for the total clouds (cyan
points) and contour measurements (green points).  We find a constant
value of $^{13}$CO versus the surface density of extinction, but find
that it underestimates $\Sigma_{\rm gas}$ by a factor of $\sim$4--5
and lies below measurements of $^{12}$CO by a factor of $\sim$5, on
average.  The difference we find between $^{13}$CO and H$_{2}$
measured by $A_{V}$, could be due to the LTE method we used to compute
$^{13}$CO masses or the assumption that there is a constant abundance
of CO relative to H$_{2}$.  \citet{heyer09} explored the properties of
galactic molecular clouds using $^{13}$CO emission and found that the
assumption of equivalent excitation temperatures for both $^{12}$CO
and $^{13}$CO in the LTE method may underestimate the true column
density of $^{13}$CO in subthermally excited regions. As the $^{13}$CO
density increases, the $J$=1--0 transition becomes thermalized, and
the column density estimates are more accurate.  Also, both
\citet{heyer09} and \citet{goldsmith08} found that if
$^{13}$CO--to--H$_{2}$ abundance variations in LTE--derived cloud
masses are not considered, they would underestimate the true column
densities by factors of 2--3.  Since we only include $^{13}$CO
emission greater than 5 times the rms noise, we are likely measuring
gas that is thermalized with little abundance variation.  Assuming a
constant abundance ratio will therefore not account for the difference
between $A_{V}$ and $^{13}$CO masses. $^{13}$CO, might therefore be a
more consistent tracer of $\Sigma_{\rm gas}$, but it may underestimate
the mass by factors of $\sim$4--5, which can be corrected for.

Variations in the CO--to--H$_{2}$ conversion factor may impact the
slope of the SFR--gas relations as measured by extragalactic
studies. Since we are resolving molecular clouds, we cannot place
constraints on gas densities lower than $\sim$50 $M_{\sun} \ \rm
pc^{-2}$, typical of spiral galaxies \citep{bigiel08}.  However if we
consider the effects of using CO as a tracer of the total gas density,
it underestimates the mass measured from $A_{V}$ by $\gtrsim$30\% for
$\Sigma_{\rm gas} \gtrsim 200 M_{\sun} \ \rm pc^{-2}$. This would
effectively shift the extragalactic observed points to the right above
200 $M_{\sun} \ \rm pc^{-2}$.  This shift would flatten the slope
slightly in the fitted \citet{k98} relation.  These small factors,
however do not explain the large discrepancy between the extragalactic
relations and the much higher SFR in the local clouds, seen both in
the whole molecular clouds and looking at the youngest YSOs as a
function of $\Sigma_{\rm gas}$.

\subsection{Do High--mass and Low--mass Star Forming Regions Behave Differently?}\label{mass}

By studying nearby molecular clouds, we can obtain the most accurate
measurement of $\Sigma_{\rm gas}$ and $\Sigma_{\rm SFR}$, but it is
regions of {\it massive} star formation that form the basis for
extragalactic studies. Massive star forming regions are the only
readily visible regions forming stars at large distances and thus are
the only probes of star formation in distant regions in the Milky Way
and external galaxies. We can measure $\Sigma_{\rm SFR}$ and
$\Sigma_{\rm gas}$ in individual massive star forming regions to see
if there is better agreement with extragalactic SFR--gas relations.

To investigate where individual regions of massive star formation fall
on the SFR--gas relation, we use data from the molecular line survey
of dense gas tracers in $>50$ massive dense ($\langle n \rangle \sim
10^{6}$ \cmv, e.g. \citet{plume97}) clumps from \citet{wu10}.  The
\citet{wu10} survey measured clump sizes, virial masses, and dense gas
surface densities using HCN $J$=1--0 as a tracer ($\Sigma_{\rm HCN}$)
at FWHM of the peak intensity. These are the sites of formation of
clusters and massive stars. The most popular tracers of massive star
formation include the ultraviolet, H$\alpha$, FIR, and singly ionized
oxygen; however, due to high extinction toward and in these regions,
we can use only the total IR luminosity to measure the SFR in these
massive clumps.

Since HCN $J$=1--0 gas has been shown to be tightly correlated with
the total IR luminosity in clumps as long as $L_{IR} > 10^{4.5}$
\lsun\ \citep{wu05}, as well as in both normal spiral and starburst
galaxies \citep{gaosol104,gaosol204}, we can use it to compare gas and
star formation from the total IR luminosity in both the Milky Way and
external galaxies.

$\Sigma_{\rm HCN}$ is calculated using the mass contained
within the FWHM size ($R_{\rm HCN}$) of the HCN gas, following the
methods used in \citet{shirley03}:

\bq\label{sigden}
\Sigma_{\rm HCN} = \pi^{-1} \times \biggl (\frac{M_{\rm vir}}{M_{\sun}} \biggr )\biggl (\frac{ R_{\rm
   HCN}}{\rm pc} \biggr)^{-2} (M_{\sun} \ \rm pc^{-2}),
\eq

where $M_{\rm vir}$ is the virial mass enclosed in the source size at
FWHM intensity.  Uncertainties in the $\Sigma_{\rm HCN}$ are computed
by adding in quadrature the errors in the FWHM size and the mass as
discussed in \citet{wu10}.

We compute the SFR for massive dense clumps following extragalactic
methods using the total infrared (IR) luminosity ($L_{\rm IR}$;
8--1000$\micron$) derived from the 4 IRAS bands. We assume the
conversion SFR$_{\rm IR}$ ($M_{\sun} \ \rm yr^{-1})\approx$ 2$\times
10^{-10}L_{\rm IR}$ ($L_{\sun}$) from \citet{kennicutt98}.
$\Sigma_{\rm SFR_{\rm IR}}$ are computed using the FWHM source sizes:

\bq\label{sfden}
\Sigma_{\rm SFR_{\rm IR}} = \pi^{-1} \times \biggl ( \frac{\rm
 SFR_{\rm IR}}{M_{\sun} \, \rm yr^{-1}}\biggr ) \biggl (\frac{R_{\rm
   HCN}}{\rm kpc} \biggr )^{-2}  (M_{\sun} \ \rm yr^{-1} \ \rm kpc^{-2}).
\eq

The uncertainties in the $\Sigma_{\rm SFR_{\rm IR}}$ density only
include the error in FWHM size and a 30\% error in SFR calibrations
using the IR. In fact, the uncertainties are larger. The SFR
calibration assumes that the observed far-IR emission is re--radiated by
dust heated by O, B, and A stars \citep{kennicutt98}. For low SFR,
heating by older stars of dust unrelated to star formation can
contaminate the SFR signal, causing an overestimate of the SFR.  This
is not a problem for the regions of massive star formation in our
Galaxy, where the heating is certainly due to the recent star
formation.  A bigger issue for the Galactic regions is that the full
$L_{\rm IR}$ seen in the extragalactic studies is not reached unless
the individual region forms enough stars to fully sample the IMF and
is old enough that the stars have reached their full luminosity. These
conditions may not be met during the time span of an individual
massive clump \citep{krumthom07, urban10}.  In particular,
\citet{urban10} have calculated the ratio of luminosity to SFR in a
simulation of a cluster forming clump. They find that $L/\rm SFR$
increases rapidly with time, but lies a factor of 3-10 below the
relation in \citet{kennicutt98} when their simulations end at times of
0.7 to 1.4 Myr. Therefore, we may expect both large variations and a
tendency to underestimate the SFR in individual
regions. Unfortunately, despite these issues, $L_{\rm IR}$ remains the
best measure of SFR available to us at present in these regions.
While it would be suspect to apply a correction factor based on the
Urban et al. simulation, increasing the SFR of the regions of massive
star formation by 0.5-1 order of magnitude would bring them into
better agreement with the highest surface density points from the
nearby clouds.

\begin{figure}
\epsscale{0.9}
\plotone{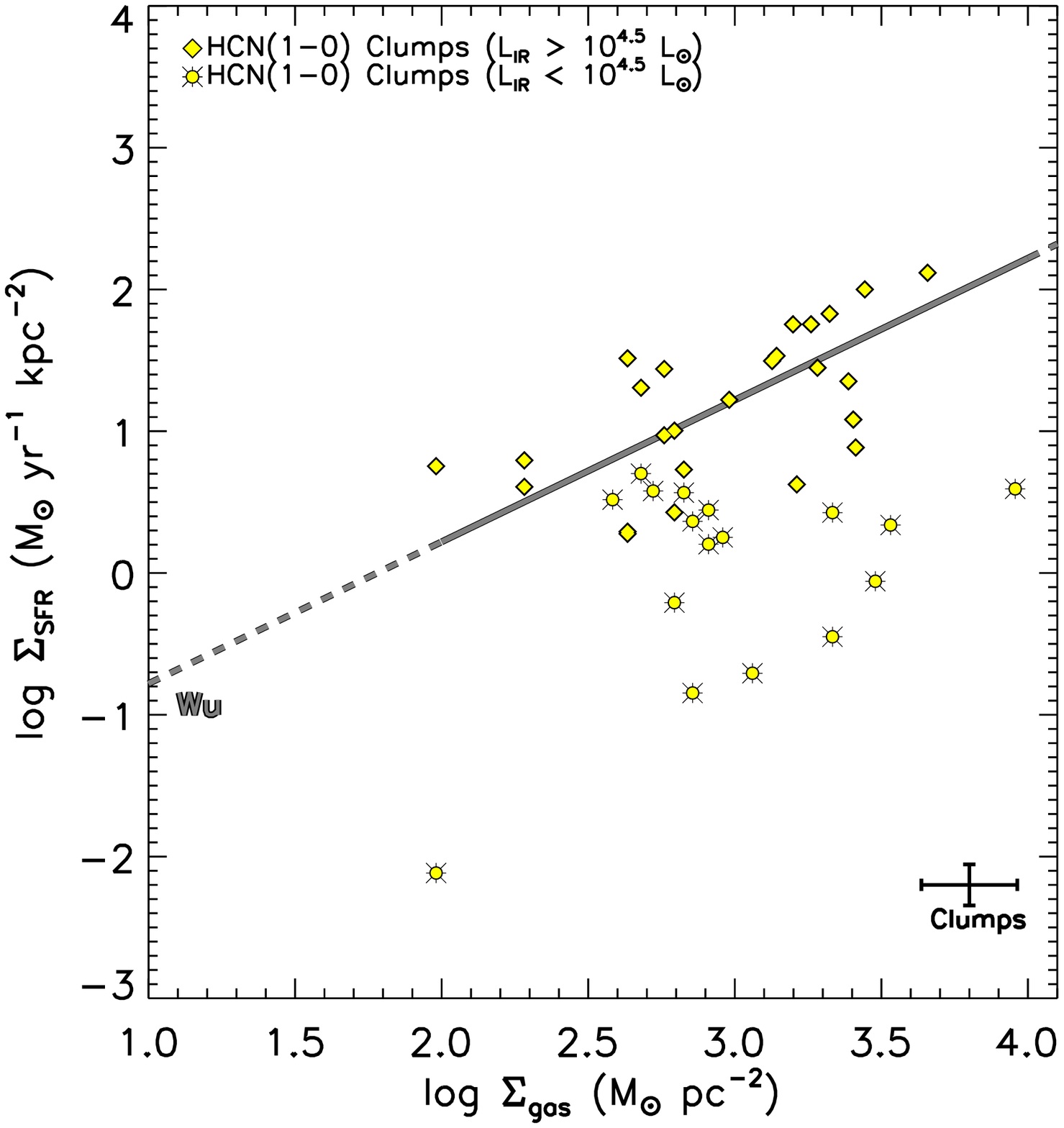}
\caption{$\Sigma_{\rm gas}$ and $\Sigma_{\rm SFR}$ are show for the
sample of massive dense clumps from the survey of \citet{wu10}.  Gas
surface densities are measured from the HCN$J$=1--0 maps and SFRs are
estimated from the total IR luminosity, using the extragalactic
prescription from \citet{k98}. The relation between SFR and dense gas
from \citet{wu05} is shown (grey solid line) and is extrapolated to
lower $\Sigma_{\rm gas}$. We make a cut at $L_{\rm IR} > 10^{4.5}
L_{\sun}$, below which the clumps are not massive enough to sample the
IMF and lie off a the linear relation (Section~\ref{mass}). }
\label{fmsf}
\end{figure}

For the sample of $\sim$50 massive dense clumps, 42 sources have
corresponding IR measurements. The resulting gas surface densities and
$\Sigma_{\rm SFR}$ for the sample of massive dense clumps are shown in
Figure~\ref{fmsf} and Table~\ref{t1msf}.  The relation between SFR and
dense gas mass, $M_{\rm dense}$(H$_{2}$), for galactic clumps can be
derived from $\langle L_{\rm IR}/L_{\rm HCN(1-0)} \rangle = 911\pm 227
(\rm K \ \rm km \ \rm s^{-1} \ \rm pc^{2})^{-1}$, and $\langle M_{\rm
dense}(H_{2})/L_{\rm HCN(1-0)} \rangle = 11\pm 2 M_{\sun} (\rm K \ \rm
km \ \rm s^{-1} \ \rm pc^{2})^{-1}$ from \citet{wu05}.  These
relations can be combined with the IR SFR conversion from above to obtain
a relation for $\Sigma_{\rm SFR_{\rm IR}}$ and $\Sigma_{\rm HCN}$:

\bq\label{sfrwu}
\Sigma_{\rm SFR_{\rm IR}} \sim 1.66\times 10^{-2}\left( \frac{\Sigma_{\rm
   HCN(1-0)}}{1 M_{\sun} \,
 \rm pc^{-2}} \right) (M_{\sun} \, \rm yr^{-1} \, \rm kpc^{-2}).
\eq

This equation is equivalent to that shown in \citet{wu05}, which is
the fit to both massive clumps and galaxies.

\begin{figure}
\epsscale{0.95}
\plotone{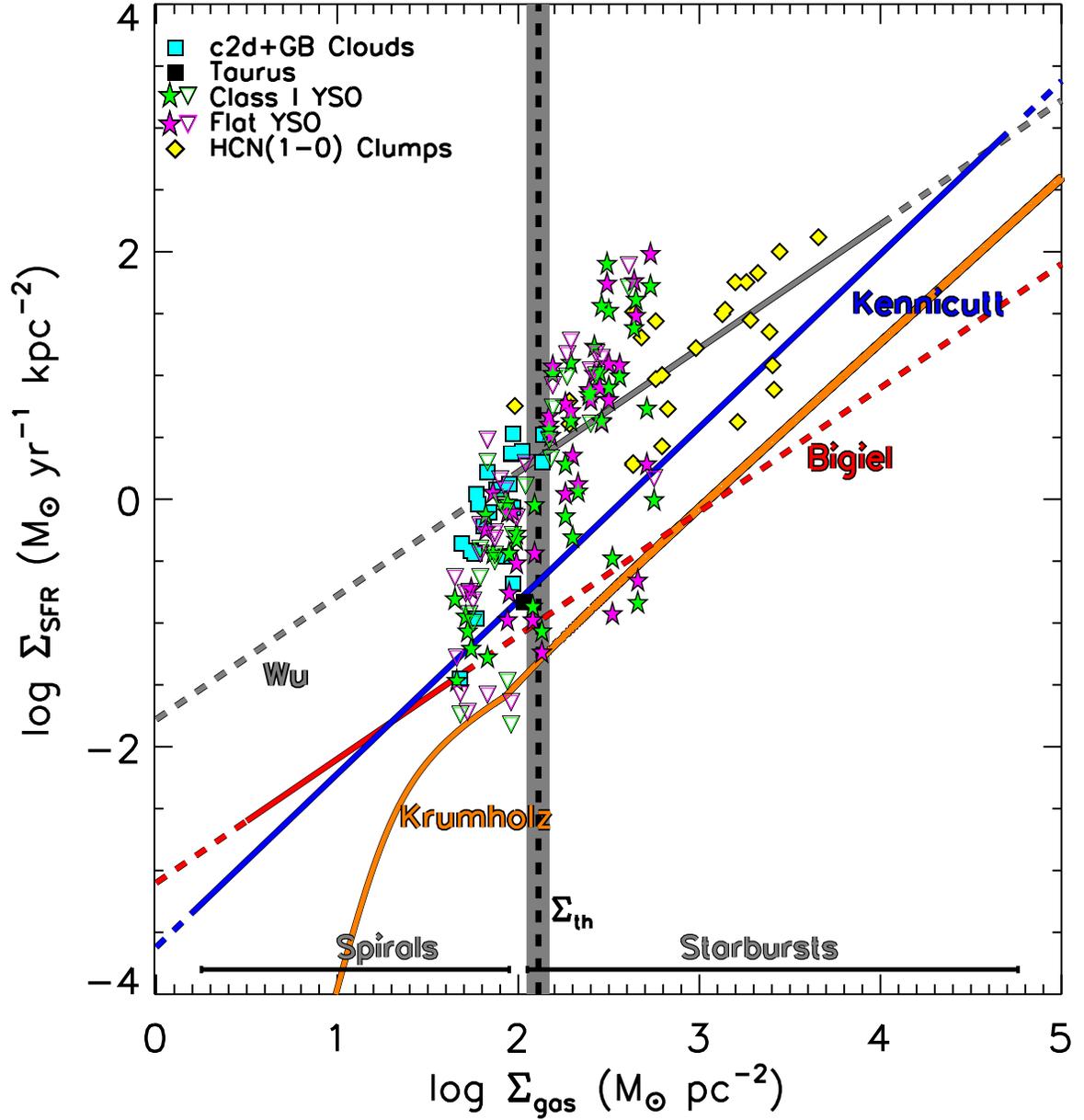}
\caption{ Comparison of Galactic total c2d and GB clouds, YSOs, and
   massive clumps to extragalactic relations.  SFR and gas surfaces
   densities for the total c2d and GB clouds (cyan squares), c2d
   Class I and Flat SED YSOs (green and magenta stars), and $L_{\rm
   IR} > 10^{4.5} L_{\sun}$ massive clumps (yellow diamonds) are
   shown. The range of gas surface densities for the spirals and
   circumnuclear starburst galaxies in the \citet{k98} sample are
   denoted by the grey horizontal lines.  The grey shaded region
   denotes the range  for  $\Sigma_{\rm th}$ of 129$\pm$14 $M_{\sun} \
   \rm pc^{-2}$.
}
\label{last}
\end{figure}

\citet{wu05} found a decline in the linear $L_{\rm}-L'_{\rm HCN(1-0)}$
correlation at $L_{\rm IR} < 10^{4.5} L_{\sun}$, where the clump is
not massive or old enough to sample the IMF. Since the majority of the
points with $L_{\rm IR} < 10^{4.5} L_{\sun}$ in Figure~\ref{fmsf} lie
off this relation, we include only massive dense clumps with $L_{\rm
IR} > 10^{4.5} L_{\sun}$ .  The resulting number of HCN$J$=1--0
sources is 25 (Table~\ref{t1msf}).  The HCN$J$=1--0 clumps are found
to range from $\sim 10^{2}$--$4.5\times 10^{3}$ $M_{\sun} \ \rm
pc^{-2}$ in $\Sigma_{\rm HCN}$ and from 2--130 $M_{\sun}\ \rm
kpc^{-2}\ \rm yr^{-1}$ in $\Sigma_{\rm SFR_{\rm IR}}$. These
$\Sigma_{\rm gas}$ and $\Sigma_{\rm SFR_{\rm IR}}$ values are similar
to those of circumnuclear starburst galaxies from \citet{k98}, which
range from $\sim10^{2}$--6$\times 10^{4}$ $M_{\sun}\ \rm pc^{-2}$ and
0.1--$9.5\times 10^{2}$ $M_{\sun}\ \rm kpc^{-2}\ \rm yr^{-1}$ (see
Figure~\ref{last}).  The average HCN $J$=1--0 clump has $\Sigma_{\rm
HCN(1-0)}$ of (1.3$\pm0.2) \times 10^{3}$ $M_{\sun} \ \rm pc^{-2}$ and
$\Sigma_{\rm SFR_{\rm IR}}$ of 28$\pm$6 $M_{\sun}\ \rm kpc^{-2}\ \rm
yr^{-1}$.

\begin{figure}
\epsscale{0.9}
\plotone{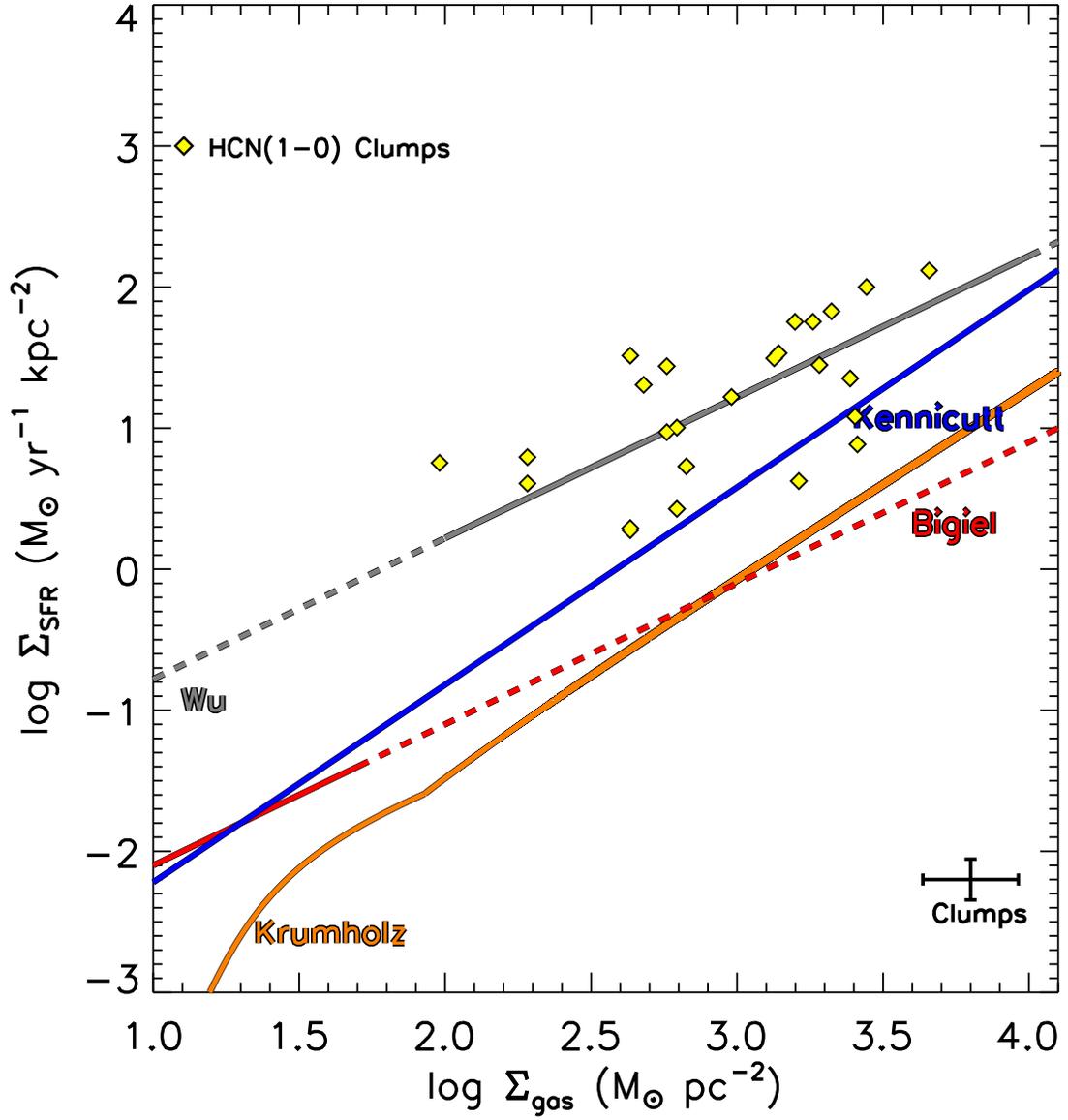}
\caption{Comparison of massive, dense clumps with $L_{\rm IR} >
10^{4.5} L_{\sun}$ to extragalactic relations (\citet{k98},
\citet{bigiel08}, and \citet{krumholz09}, blue, red, and orange lines,
respectively). The relation between SFR and dense gas from
\citet{wu05} is also shown (grey solid line).
}
\label{fmclump}
\end{figure}

We also compare the relation we find for the massive dense clumps to
known extragalactic relations in Figure~\ref{fmclump}.  In this figure
we compare to the H$_{2}$ gas surface density relation from
\citet{bigiel08} and the total ({\h1}+H$_{2}$) gas surface density
relation from \citet{k98}. We find that most of the points for the HCN
$J$=1--0 line lie above both the \citet{bigiel08} and \citet{k98}
extragalactic relations with the average clump lying a factor of
$\sim$~5--20 above the \citet{k98} and \citet{bigiel08} relations,
respectively.

In Figure~\ref{sigth} we plot the ratio of $\Sigma_{\rm
SFR}/\Sigma_{\rm gas}$ versus $\Sigma_{\rm gas}$ for c2d and GB
clouds, YSOs, and massive clumps.  We find a steep decline in
$\Sigma_{\rm SFR}$ and $\Sigma_{\rm SFR}/\Sigma_{\rm gas}$ at around
$\sim$100--200 $M_{\sun}\ \rm pc^{-2}$.  We identify this steep change
in $\Sigma_{\rm SFR}$ over $\sim$100--200 $M_{\sun}\ \rm pc^{-2}$ as a
star forming threshold ($\Sigma_{\rm th}$) between regions actively
forming stars and those that are forming few or no low--mass stars.

In Figure~\ref{last} we show points for the massive dense clumps, c2d
and GB clouds, the youngest YSOs, and both the \citet{wu05} and
extragalactic relations.  We also show the range of gas surface
densities for spiral and circumnuclear starburst galaxies from the
sample of \citet{k98}. $\Sigma_{\rm gas, con}$ for Class I and Flat
SED YSOs lie intermediate between the regions where spiral galaxies
and starburst galaxies are found on the \citet{k98} relation. At
$\Sigma_{\rm gas}>$ $\Sigma_{\rm th}$, the youngest Class I and Flat
SED YSOs overlap with the massive clumps (Figure~\ref{last}).
Therefore, high--mass and low--mass star forming regions behave
similarly in the $\Sigma_{\rm SFR}$--$\Sigma_{\rm gas}$ plane. The
difference between extragalactic relations and c2d and GB clouds is
not caused by the lack of massive stars in the local clouds.  Also,
the overlap with the massive clumps in Figure~\ref{last} suggests that
$L_{\rm IR}$ provides a reasonable SFR indicator, as long as it
exceeds $10^{4.5}$ \lsun, though an upward correction would produce
better agreement.

A steep increase and possible leveling off in $\Sigma_{\rm SFR}$ at a
threshold $\Sigma_{\rm th}\sim$100--200 $M_{\sun}\ \rm pc^{-2}$ is
seen in both Figures~\ref{sigth} and~\ref{last}. We further constrain
this steep increase and the possibility of $\Sigma_{\rm SFR}$
flattening at $\Sigma_{\rm gas} > \Sigma_{\rm th}$, by approximating
it as broken power law with a steep rise that levels off in
Section~\ref{sfth}.

\subsubsection{Star Formation Threshold}\label{sfth}

\begin{figure}
\epsscale{0.9}
\plotone{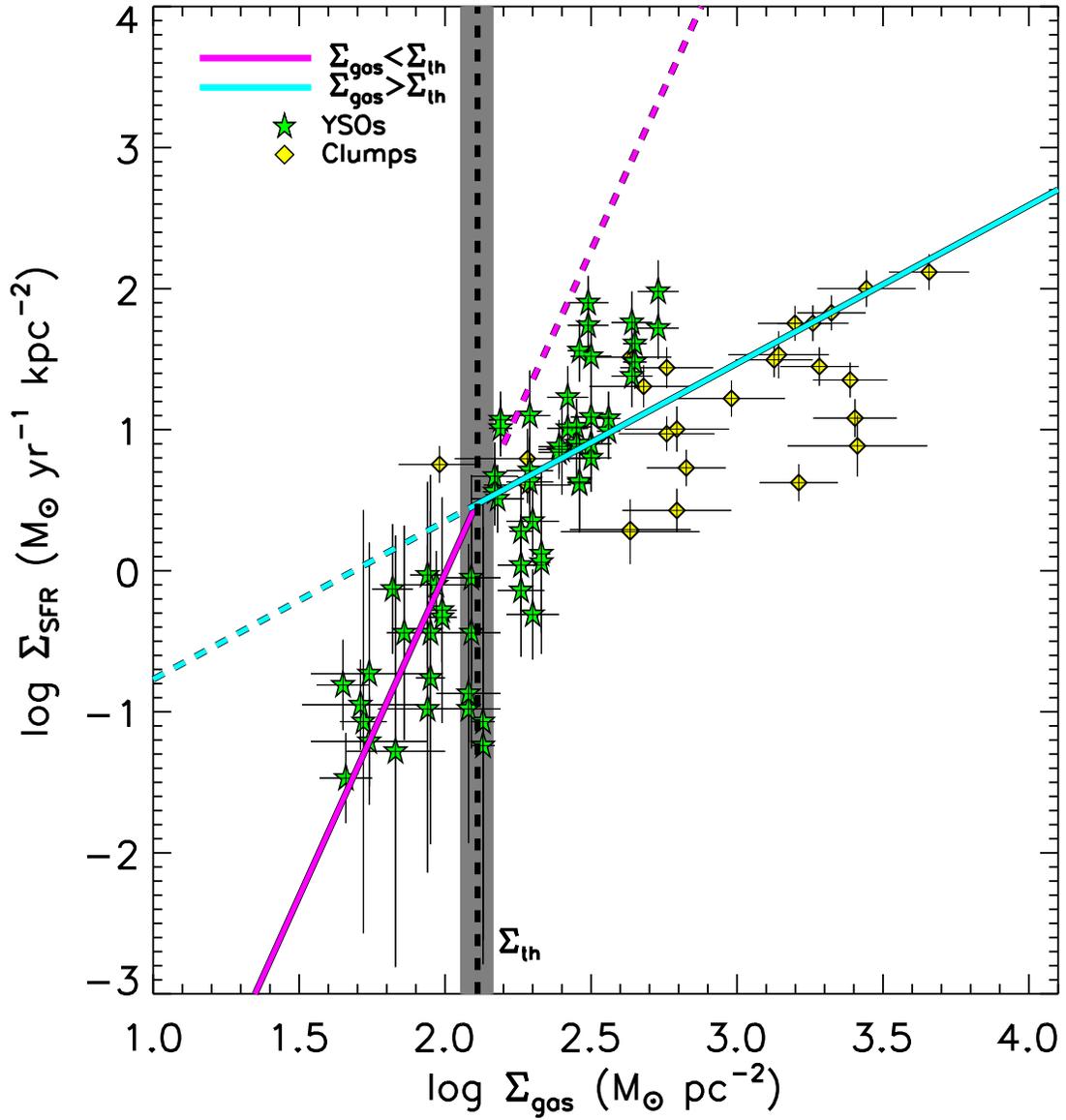}
\caption{ We fit Class I and Flat SED YSOs (green stars) and massive
clumps (yellow diamonds) to a broken power law (Section~\ref{mass})
and obtain an estimate for the star forming threshold, $\Sigma_{\rm
th}$, of 129$\pm$14 $M_{\sun}$ pc$^{-2}$ (grey shaded region). The
slope changes from 4.6 below $\Sigma_{\rm th}$ to 1.1 above
$\Sigma_{\rm th}$.}
\label{bplfit}
\end{figure}

In order to determine a robust estimate of $\Sigma_{\rm th}$, we fit
the data using two models: a single power law ($y=Nx + A$, where
$y=$log$_{10} \ \Sigma_{\rm SFR}$; x= log$_{10} \ \Sigma_{\rm gas}$
and a broken power law ($y_{1}=N_{1}x + A_{1}$; $y_{2}=N_{2}x +
A_{2}$).  We first fit Class I and Flat SED YSO points ($\Sigma_{\rm
SFR, con}, \Sigma_{\rm gas, con}$) and massive clumps ($\Sigma_{\rm
SFR_{\rm IR}}, \Sigma_{\rm HCN(1-0)}$) to a single power law.  We do
not include upper limits for YSOs or points for Sco and Cep clouds,
which are co--added separate cloud regions and only provide a rough
estimate of $\Sigma_{\rm SFR}$ and $\Sigma_{\rm gas}$.  The single
power law fit yields $N=1.57\pm0.09$ and $A= -3.0\pm0.2$, and a
reduced chi-square ($\chi^{2}_{r}$) of 3.7 (84 dof).  We fit the data
for both YSOs and massive clumps to a broken power law for the range
of $\Sigma_{\rm gas}=$50--250 $M_{\sun}\rm pc^{2}$.  We minimize the
total $\chi^{2}$ for the two segments of the broken power law using a
simplex routine, which yields best fit parameters: $N_{1}=
4.58\pm0.5$, $A_{1}=-9.18\pm0.9$, $N_{2}=1.12\pm0.07$,
$A_{2}=-1.89\pm0.2$ with a $\chi^{2}_{r}$ of 3.04 (82 dof).  We
attribute the large $\chi_{r}^{2}$ to the scatter and large errors in
the data, but since the $\chi_{r}^{2}$ is $\sim$18\% lower for the
broken power law compared to the single power law, we take it to be
the best fit model.  Equating the broken power law fits
($y_{1}=y_{2}$), we obtain a power law break at $\Sigma_{\rm th}$=
129.2 $M_{\sun}\ \rm pc^{2}$ ($A_{V}=8.6$) with a statistical 1--sigma
deviation in $\chi^{2}$ of $\pm14$ $M_{\sun}\ \rm pc^{2}$ giving a
range in $\Sigma_{\rm th}$ of $\sim$115--143$M_{\sun}\ \rm pc^{2}$.
Figure~\ref{bplfit} shows the broken power law fit (cyan and magenta
lines), $\Sigma_{\rm th}$, and the 1--sigma statistical range of
$\Sigma_{\rm th}$ (dashed black vertical line and grey shaded region,
respectively). The slope of the broken power law changes from a steep
relation at $\Sigma_{\rm gas} < \Sigma_{\rm th}$ (slope of $\sim$4.6)
to linear relation (slope of $\sim$1.1) at $\Sigma_{\rm gas} >
\Sigma_{\rm th}$. We note however, that variations in cloud distances
will change this threshold slightly. One example is for Ser, which has
a recent distance estimate by \citet{dzib10} of 415 pc and another
estimate by \citet{strai96} of 260 pc (used in this paper).  If we use
the larger distance of 415 pc, this would change our star formation
threshold slightly to 126$\pm$12 $M_{\sun}\ \rm pc^{2}$ ($A_{V}=8.4$).

This star forming threshold we find is in agreement with the threshold
found in a study of local molecular clouds by Lada, Lombardi, \& Alves
(2010, submitted) at $A_{V}\sim7$ or 116 $M_{\sun}\ \rm
pc^{2}$. \citet{enoch07} also found extinction thresholds for dense
cores found in Bolocam 1.1mm maps in the Perseus, Serpens, and
Ophiuchus clouds of $A_{V}\sim$8, 15, and 23, respectively ($\sim$120,
225, and 350 $M_{\sun}\ \rm pc^{2}$), with a low probability of
finding cores below these thresholds. \citet{onishi98} surveyed Taurus
in C$^{18}$O and found a star forming column density threshold of
8$\times 10^{21}\rm \ cm^{-2}$, which corresponds to a gas surface
density of 128.1 $M_{\sun}\ \rm pc^{2}$ ($A_{V}$=8.5). Similarly, both
\citet{johnstone04} and \citet{andre10} find thresholds of $A_{V}\sim$10 (150 $M_{\sun}\ \rm pc^{2}$).

\citet{mouschovias76} proposed the idea of a physical column density
threshold corresponding to the central surface density above which the
interstellar magnetic field cannot support the gas from self
gravitational collapse.  This was later modified by \citet{mckee89}
who considered the local ionization states owing to UV
radiation. \citet{mouschovias76} predicted that when clumps combine to
form a large cloud complex, there exists a natural surface density
threshold ($\Sigma_{\rm crit}$) for a given magnetic field:

\bq\label{bden} \Sigma_{\rm crit} > \biggl( \frac{80}{M_{\sun} \,\rm
 pc^{2}}\biggr ) \times \biggl( \frac{B}{30 \, \mu \rm G}\biggr ).
 \eq

The total mean strength of the line--of--sight magnetic field
 ($B_{los}$) measured in molecular clouds ($n\sim 10^{3}$--$10^{4}$
 cm$^{3}$) is $\sim$10--20$\mu\rm G$ \citep{crutcher99,troland08}.
 Since statistically $B_{los} \approx \frac{1}{2}B_{tot}$
 \citep{heiles05}, the corresponding total magnetic field, $B_{tot}$,
 is $\sim$20--40$\mu\rm G$.  Using equation~\ref{bden}, the
 corresponding $\Sigma_{\rm crit} >$50--110 $M_{\sun}\ \rm pc^{2}$.
 A similar idea of a threshold at a particular extinction was
 predicted by \citet{mckee89} for photoionization--regulated star
 formation.  This model predicts that the rate of star formation is
 controlled by ambipolar diffusion and therefore depends on the
 ionization levels in the cloud.  Star formation in a ``standard''
 ionization case will occur at $A_{V}\gtrsim$4--8 mag, which
 translates into a $\Sigma_{\rm crit} \gtrsim$60--120 $M_{\sun}\ \rm
 pc^{2}$.  Both of these predictions for a critical density of star
 formation are similar to $\Sigma_{\rm th}=$129$\pm$14 $M_{\sun}\
 \rm pc^{2}$. We note, however, that both these models are for parts
 of clouds that are in a ``quasi--static'' or turbulence-supported
 state.  Alternatively, these parts of clouds may never
 become bound and are transient \citep{vazsem09}.  In this picture,
 the threshold would correspond to parts of molecular clouds that
 become gravitationally bound and form stars.

\subsection{Does the Lack of Resolution in Extragalactic Studies
 Explain the Discrepancy in $\Sigma_{\rm SFR}$?}\label{res}

The third possibility is based on the fact that the extragalactic
relations are averaging over large scales which do not resolve the
regions where stars are forming.  Current ``spatially resolved"
extragalactic measurements are still limited to scales of $\sim$0.2-2
kpc \citep{kennicutt07, bigiel08, blanc09}; therefore, we cannot
directly measure extragalactic SFRs on scales of galactic star forming
regions.  In any given spatially resolved extragalactic measurement of
$\Sigma_{\rm SFR}$ and $\Sigma_{\rm gas}$, the beam will contain a
fraction of diffuse gas that does not trace star formation and a
fraction of dense, star forming gas ($f_{\rm dense}$).  As discussed
in Section~\ref{comap}, the dense gas that is forming stars is not
well traced by CO so there will be an excess of non--star forming gas
in each beam measurement.  A local example of this is a study of the
Taurus molecular cloud; \citet{goldsmith08} found that 33\% of diffuse
$^{12}$CO is contained in regions not associated with $^{13}$CO, which
is a more reliable tracer of dense, star forming gas
(Section~\ref{comap}).  Even in the regions with $A_{V} >2$ studied by
the c2d and GB projects, star formation is highly concentrated to
regions of high extinction (e.g., Figure~\ref{cont}). Extragalactic
studies averaging over hundreds of pc scales would include this
diffuse gas, causing an increase in the amount of CO flux that is
being counted as star forming gas.  In order to better understand
approximately how much gas is forming stars at present, a measurement
of the fraction of gas that contains YSOs over a larger area on kpc
scales in the Galaxy is needed.

\citet{lada03} proposed the idea that clusters of stars which form in
clumps located in giant molecular clouds are the fundamental building
blocks of galaxies.  The rate at which these stellar clusters form is
set by the mechanisms that enable these clumps to condense out of
their low density parent cloud.  A similar idea was explored by
\citet{wu05}, who proposed that there is a basic unit of clustered
star formation with the following typical properties: $L_{\rm IR} >
10^{5} L_{\sun}$, $R_{\rm dense}\sim$0.5 pc, and $M_{\rm
vir}\sim$300--1000 $M_{\sun}$.  As more of these basic units are
contained in a galaxy, the SFR increases linearly. This linear
correlation between SFR$_{\rm IR}$ and the mass of dense gas ($M_{\rm
dense}$) from HCN $J$=1--0 was seen in a sample of both spiral and
luminous or ultra--luminous IR galaxies (LIRGS and ULIRGS)
\citep{gaosol204}, and \citet{wu05} showed that the same relation fit
the Galactic massive dense clumps. It is therefore the dense gas
tracers, such as HCN, that directly probe the volume of gas from which
stars form in dense clumps and produce the star formation in external
galaxies.

If a linear relation between dense gas and the SFR is assumed at all
$\Sigma_{\rm gas}$ and $\Sigma_{\rm SFR}$ densities, how can we
explain the non--linear behavior of the Kennicutt--Schmidt SFR--gas relation?
Let us suppose that the underlying star formation relation is what we
actually observe in regions forming massive stars: a threshold around
129 \msun\ pc$^{-2}$ and a roughly linear relation between dense gas
and star formation above that threshold:

\bq\label{sigdense}
\Sigma_{\rm SFR} \propto \Sigma_{\rm dense}.
\eq

Let us also assume, the Kennicutt--Schmidt relation for the gas surface
density averaged over large scales ($\mean{\Sigma_{\rm gas}}$):

\bq\label{ksrelation}
\Sigma_{\rm SFR} \propto \mean{\Sigma_{\rm gas}}^{1.4}.
\eq

 Then,  the fraction of gas above the threshold ($f_{\rm dense}$) would have to scale with mean surface density of all gas:
\bq\label{fdenseeq}
\begin{split}
f_{\rm dense} = \Sigma_{\rm dense}/ \mean{\Sigma_{\rm gas}}\\
\propto \Sigma_{\rm SFR}/ \mean{\Sigma_{\rm gas}}\\
\propto \mean{\Sigma_{\rm gas}}^{0.4}.
\end{split}
\eq

Near the $\Sigma_{\rm th}$ of $\sim$129 $M_{\sun} \ \rm pc^{-2}$, the
average $\Sigma_{\rm SFR}$ measured on small scales is about $\sim$3
$M_{\sun} \ \rm yr^{-1}\ \rm kpc^{-2}$ (taking the average between
Class I and Flat SED sources) and $f_{\rm dense}$ is about 1/40. When
$\mean{\Sigma_{\rm gas}} \sim 300 \Sigma_{\rm th}$, $f_{\rm dense}
\sim 1$.  At this point, all the gas is dense enough to form stars and
star formation is most efficient, creating a maximal starburst.  This
is also where the dense gas \citep{wu05} and CO \citep{kennicutt98}
relations cross (see Figure~\ref{sigth}).  Above $\sim$300$\Sigma_{\rm
th}$, the only way to increase star formation efficiency is to make it
more efficient even in the dense gas. This is possible because even in
dense gas, the star formation rate per free-fall time is less than
unity \citep{zuckerman74,krumtan07}.

Is there local evidence for a preponderance of gas below the
threshold?  Complete maps of CO for the local 0.5 kpc are not readily
available, but we can measure the mass below and the mass above the
threshold of $A_{V} = 8$ for the 16 clouds with Spitzer coverage down
to $A_{V} = 2$. The ratio of total mass lying below the threshold to
the total mass above the threshold is 4.6. The massive star forming
region, Orion, also has a similar ratio of 5.1 for the mass below over
the mass above $A_{V} = 8$ (M. Heyer, unpublished data). A few clouds
have been mapped to still lower levels and a factor of two more mass
is found in Taurus, for example \citet{goldsmith08}. Alternatively, if
we assume Orion to contain the largest reservoir of molecular material
within 0.5 kpc, we can derive the ratio of mass from $^{12}$CO and
$A_{V}$ maps to get 6.4 (M. Heyer, unpublished data). Taken together,
these factors make it plausible that there is 10 times more molecular
mass than mass above the threshold.  Furthermore, most of the gas
within 0.5 kpc is atomic. If that is included, the predictions of the
extragalactic Kennicutt--Schmidt relation for total gas agree
reasonably with the local star formation rate surface density (see
\citet{evans08} and references therein).

Finally, if the underlying star formation law in the dense gas is
linear, then the arguments invoking the density dependence of the
free-fall time to get a power of 1.5 (see Section~\ref{intro}) are
specious. In fact, the idea of a single free-fall time for a molecular
cloud with an enormous range of densities is highly dubious in the
first place.

\section{Summary}\label{summary}
\begin{figure}
\epsscale{0.9}
\plotone{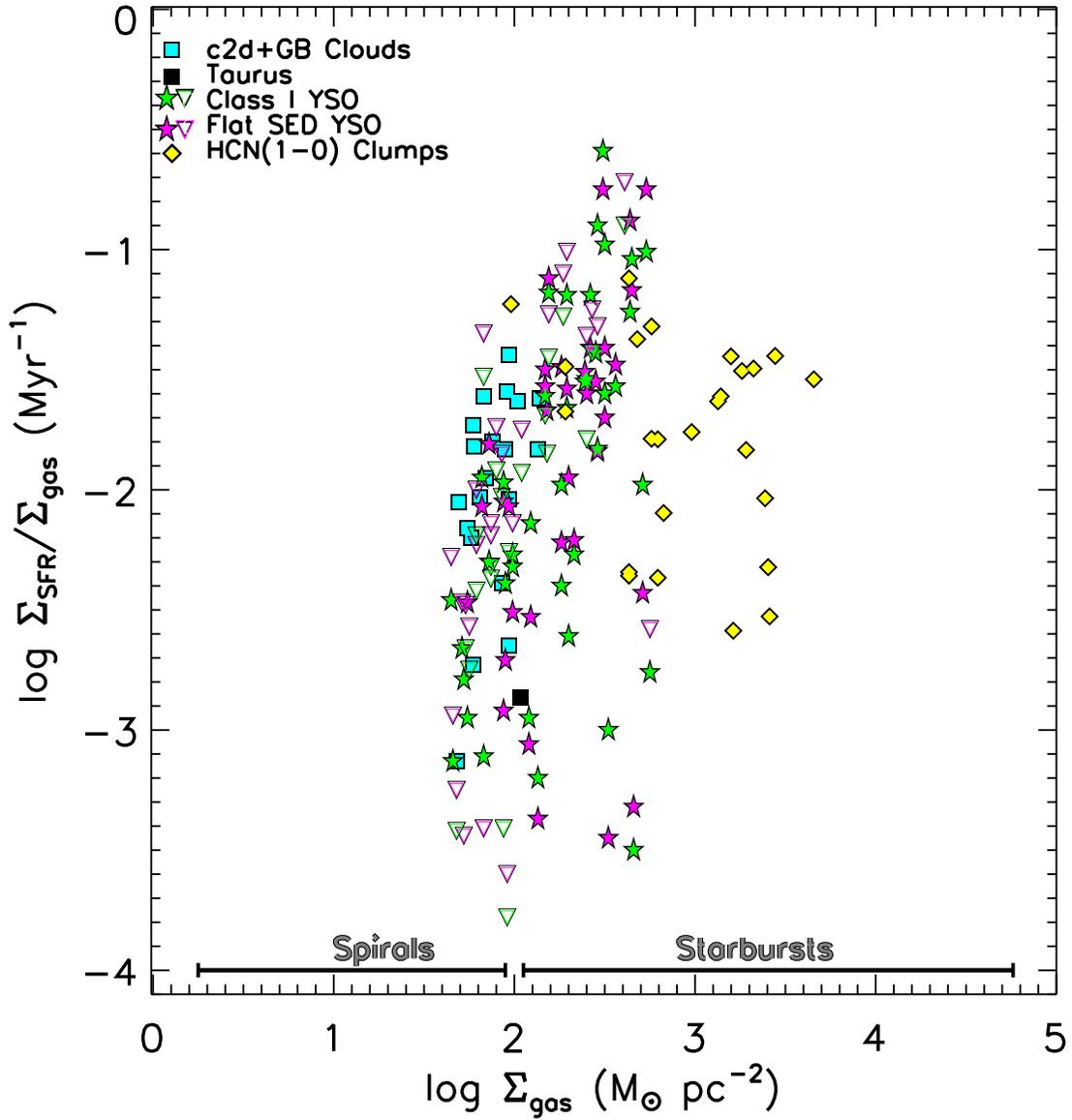}
\caption{The ratio of $\Sigma_{\rm SFR}$ and $\Sigma_{\rm gas}$
compared to $\Sigma_{\rm gas}$ for low and high--mass star forming
regions.  We find a steep fall off in $\Sigma_{\rm SFR}/\Sigma_{\rm
gas}$ in the range of $\Sigma_{\rm gas}\sim$100--200 $M_{\sun}$
pc$^{-2}$.  We denote this steep fall off as a star forming threshold,
$\Sigma_{\rm th}$, between active star forming regions and inactive
regions.}
\label{sigth}
\end{figure}

We investigate the relation between star formation rate (SFR) and gas
surface densities in a sample of young stellar objects (YSOs) and
massive dense clumps.  Our YSO sample consists of objects located in
20 large molecular clouds from the \spitzer \ cores to disks (c2d) and
Gould's Belt (GB) surveys.  We estimate the $\Sigma_{\rm gas}$ in the
c2d and GB clouds from $A_{V}$ maps and $\Sigma_{\rm SFR}$ from the
number of YSOs, assuming a mean mass and star formation timescale for
each source. We also divide the clouds into evenly spaced contour levels
of $A_{V}$. In each contour interval, we measure the $\Sigma_{\rm gas,
con}$ and estimate the $\Sigma_{\rm SFR}$ by counting only Class I and
Flat SED YSOs which have not yet migrated from their birthplace.  We
use $^{12}$CO and $^{13}$CO gas maps of the Perseus and Ophiuchus
clouds from the COMPLETE survey to estimate $\Sigma_{\rm H_{2}}$
densities and compare to measurements from $A_{V}$ maps.  We also
compare the c2d and GB low--mass star forming regions to a sample of
massive star forming clumps from \citet{wu10}. We derive SFRs from
the total IR luminosity and use HCN gas maps to estimate SFR surface
densities ($\Sigma_{\rm SFR_{\rm IR}}$) and gas surface densities
($\Sigma_{\rm HCN}$) for the massive clumps. Our results are as
follows:

\begin{itemize}

\item
The c2d and GB clouds lie above the extragalactic SFR-gas
relations (e.g., Kennicutt--Schmidt Law) by factors of 9--17.  We
compare the total cloud points to the theoretical prediction of
\citet{krumholz09} for galactic metallicity and a clumping factor of
1, corresponding to scales of 100 pc, and find the clouds to lie above
this prediction by a factor of $\sim$40.

\item
We perform a follow up survey of suspicious YSOs (MISFITS) at the CSO
using the \hcop$J$=3--2 line transition as a dense gas tracer. We
choose the youngest YSOs (Class I and Flat SED) that have not yet had
time to migrate from their birthplace.  These sources are spatially
positioned at low extinction levels, most are not clustered, and most
lie outside the $A_{V}$ peaks.  In this paper, we present results for
a total of 98 sources, including 45 Flat SED and 53 Class I YSOs
(detailed results from the full survey will be published in a later
paper).  We find that 74\% or 73 out of the 98 MISFITS observed to
date are not detected in \hcop \, which indicates that they do not
have a dense envelope of gas, and could be either later class YSOs or
background galaxies. These are a small fraction of the total number of
YSOs in the sample, but they could bias the statistics upward at low
$\Sigma_{\rm gas}$.

\item
 We divide the c2d and GB clouds into contours using evenly spaced
 intervals of $A_{V}$ (Section~\ref{sfysot}).  We count only the
 youngest YSOs, removing any Class I or Flat SED YSOs (MISFITS) that
 are not detected in \hcop (Sections~\ref{misfit} and~\ref{ysocont}).
 We find that the observed extragalactic relations \citep{k98,
 bigiel08} under--predict the average $\Sigma_{\rm SFR}$ of $\sim$9.7
 $M_{\sun} \ \rm yr^{-1}\ \rm kpc^{-2}$ by factors of $\sim$21--54 and
 that our data lie above the theoretical relation \citep{krumholz09}
 by $\sim$2 orders of magnitude.

\item
 We compare $\Sigma_{\rm gas}$ calculated from $A_{V}$ maps to
 $\Sigma_{\rm H_{2}}$ estimated from $^{12}$CO in
 Section~\ref{comap}. We find that the mass estimated from $^{12}$CO
 may underestimate the $\Sigma_{\rm gas}$ at $\Sigma_{\rm gas}
 \gtrsim$200 $M_{\sun} \ \rm pc^{-2}$ by $>$30\% (Figure~\ref{sigth}).
 If the $\Sigma_{\rm H_{2}}$ from $^{12}$CO underestimates the H$_{2}$
 mass at $\Sigma_{\rm gas} \gtrsim$200 $M_{\sun} \ \rm pc^{-2}$, then
 this would effectively shift the extragalactic observed data to the
 right above this threshold, flattening the slope in the \citet{k98}
 relation. However, this small change is not enough to account for the
 discrepancy between Galactic and extragalactic measurements.

\item
We also compare $\Sigma_{\rm H_{2}}$ from $^{13}$CO maps to
 $\Sigma_{\rm gas}$ from $A_{V}$ maps. If $^{13}$CO traces the mass we
 find using extinction maps, we would expect the ratio of
 $^{13}$CO/$A_{V}$ mass to be of order unity. However, we find the
 mass estimated from $^{13}$CO to be lower than the $A_{V}$ mass by
 factors of $\sim$4--5 (Section~\ref{comap}).

\item
We find a steep decrease in $\Sigma_{\rm SFR}/\Sigma_{\rm gas}$
(Figure~\ref{sigth}) and denote this as a star formation threshold
($\Sigma_{\rm th}$).  In order to determine $\Sigma_{\rm th}$, we fit
a single power law and broken power law models to data for Class I and
Flat SED YSOs and massive clumps.  We find the best fit to the
SFR--gas relation between YSOs and clumps to be a broken power law
(Section~\ref{sfth}) with a break $\Sigma_{\rm th}$ = 129$\pm$14
$M_{\sun}$ pc$^{-2}$.  We find a steep relation at $\Sigma_{\rm gas} <
\Sigma_{\rm th}$ (slope of $\sim$4.6) and a linear relation at
$\Sigma_{\rm gas} > \Sigma_{\rm th}$ with a slope of $\sim$1.1
(Section~\ref{sfth}).

\item
Since the c2d and GB clouds are forming low--mass stars, and
extragalactic studies are only able to use tracers that measure the
light coming from massive stars, the two star forming regimes might
behave differently, accounting for the large difference we
measure. However, we find that both high and low--mass star forming
regions in the Galaxy follow roughly the same linear relation above
$\Sigma_{\rm th}$ (Section~\ref{mass}).

\item
A contributing factor to the difference seen between Milky Way clouds
and extragalactic measurements both on disk--averaged and spatially
resolved scales is that extragalactic measurements average over large
scales.  These measurements include both star forming gas and gas that
is not dense enough to form stars.

\item
 Assuming the Kennicutt-Schmidt relation and that the fundamental
 correlation between $\Sigma_{\rm SFR}$ and the dense gas
 ($\Sigma_{\rm dense}$) is linear, then the fraction of dense
 star-forming gas is proportional to $\mean{\Sigma_{\rm gas}}^{0.4}$.
 When $\mean{\Sigma_{\rm gas}}$ reaches $\sim$300$\Sigma_{\rm th}$,
 the fraction of dense gas is $\sim$1, creating a maximal starburst.

\end{itemize}

\acknowledgments The authors thank Mark Krumholz, Charles Lada,
Guillermo Blanc, Miranda Dunham, Amanda Bayless, and Jaime Pineda for
informative discussions. A.H. and N.J.E. acknowledge support for this
work, part of the Spitzer Legacy Science Program, provided by NASA
through contract 1288664 issued by the Jet Propulsion Laboratory,
California Institute of Technology, under NASA contract 1407 and from
NSF Grant AST--0607793 to the University of Texas at Austin, and the
State of Texas.  L.A. and T.H. are supported by the Gould's Belt
Spitzer Legacy grant 1298236. M.H is supported by NSF grant
AST--0832222.



\begin{landscape}
\begin{deluxetable}{lrrrccrrrrc}
\tabletypesize{\scriptsize}
\tablewidth{0pt}
\tablecaption{\label{t1} Measured Quantities for Clouds.}
\tablehead{
\colhead{Cloud} &
\colhead{$N_{\rm YSOs,tot}$}&
\colhead{$N_{\rm YSOs,I}$}&
\colhead{$N_{\rm YSOs,F}$}&
\colhead{Distance}&
\colhead{$\Omega$}&
\colhead{$A_{\rm cloud}$} &
\colhead{$M_{\rm gas, cloud}$}&
\colhead{$\Sigma_{\rm gas, cloud}$}&
\colhead{SFR}&
\colhead{$\Sigma_{\rm SFR}$}\\
\colhead{}&
\colhead{}&
\colhead{}&
\colhead{}&
\colhead{(pc)}&
\colhead{(deg$^{2}$)}&
\colhead{(pc$^{2}$)}&
\colhead{($M_{\sun}$)}&
\colhead{($M_{\sun}$}&
\colhead{($M_{\sun}$}&
\colhead{($M_{\sun}$ yr$^{-1}$}\\
\colhead{}&
\colhead{}&
\colhead{}&
\colhead{}&
\colhead{}&
\colhead{}&
\colhead{}&
\colhead{}&
\colhead{pc$^{-2}$)}&
\colhead{Myr$^{-1}$)}&
\colhead{kpc$^{-2}$)}\\
\colhead{(1)} &
\colhead{(2)} &
\colhead{(3)} &
\colhead{(4)} &
\colhead{(5)} &
\colhead{(6)} &
\colhead{(7)} &
\colhead{(8)} &
\colhead{(9)} &
\colhead{(10)} &
\colhead{(11)} \\
}
\startdata
Cha II &  24 &  0 &  2 &  178$\pm$18 &  1.0 &  9.9$\pm$2.0 &  637.4$\pm$296.3 &  64.3$\pm$26 &  6.0$\pm$3.2 &  0.61$\pm$0.35\\
Lup I &  13 &  2 &  1 &  150$\pm$20 &  1.3 &  8.9$\pm$2.4 &  512.5$\pm$308.2 &  57.9$\pm$31 &  3.2$\pm$1.8 &  0.37$\pm$0.22\\
Lup III &  68 &  2 &  6 &  200$\pm$20 &  1.3 &  15.4$\pm$3.1 &  912.1$\pm$517.3 &  59.1$\pm$31 &  17.0$\pm$9.2 &  1.10$\pm$0.63\\
Lup IV &  12 &  1 &  0 &  150$\pm$20 &  0.4 &  2.5$\pm$0.7 &  189.3$\pm$95.5 &  75.1$\pm$32 &  3.0$\pm$1.6 &  1.19$\pm$0.72\\
Oph &  290 &  27 &  44 &  125$\pm$25 &  6.2 &  29.6$\pm$11.8 &  3115.3$\pm$1754.3 &  105.4$\pm$41 &  72.5$\pm$39.0 &  2.45$\pm$1.65\\
Per &  385 &  76 &  35 &  250$\pm$50 &  3.8 &  73.2$\pm$29.3 &  6585.3$\pm$3557.1 &  90.0$\pm$32 &  96.2$\pm$51.8 &  1.32$\pm$0.88\\
Ser &  224 &  31 &  21 &  260$\pm$10 &  0.8 &  17.0$\pm$1.3 &  2336.7$\pm$640.2 &  137.3$\pm$36 &  56.0$\pm$30.2 &  3.29$\pm$1.79\\
AurN &  2 &  1 &  0 &  300$\pm$30 &  0.1 &  2.4$\pm$0.5 &  223.9$\pm$51.9 &  92.8$\pm$10 &  0.5$\pm$0.3 &  0.21$\pm$0.12\\
Aur &  171 &  43 &  24 &  300$\pm$30 &  1.8 &  50.0$\pm$10.0 &  4617.5$\pm$1072.7 &  92.4$\pm$10 &  42.8$\pm$23.0 &  0.86$\pm$0.49\\
Cep &  118 &  30 &  10 &  300$\pm$30 &  1.4 &  38.0$\pm$7.6 &  2610.3$\pm$168.5 &  68.7$\pm$17 &  29.5$\pm$15.9 &  0.78$\pm$0.45\\
Cha III &  4 &  1 &  0 &  200$\pm$20 &  2.3 &  28.0$\pm$5.6 &  1326.0$\pm$386.2 &  47.4$\pm$10 &  1.0$\pm$0.5 &  0.04$\pm$0.02\\
Cha I &  89 &  10 &  12 &  200$\pm$20 &  0.8 &  9.4$\pm$1.9 &  857.3$\pm$206.3 &  91.1$\pm$12 &  22.2$\pm$12.0 &  2.36$\pm$1.36\\
CrA &  41 &  7 &  3 &  130$\pm$25 &  0.6 &  3.0$\pm$1.2 &  279.2$\pm$114.0 &  92.3$\pm$12 &  10.2$\pm$5.5 &  3.39$\pm$2.24\\
IC5146E &  93 &  13 &  9 &  950$\pm$80 &  0.2 &  61.4$\pm$10.3 &  3365.2$\pm$872.9 &  54.8$\pm$10 &  23.2$\pm$12.5 &  0.38$\pm$0.21\\
IC5146NW &  38 &  16 &  3 &  950$\pm$80 &  0.3 &  87.6$\pm$14.8 &  5178.1$\pm$1257.3 &  59.1$\pm$10 &  9.5$\pm$5.1 &  0.11$\pm$0.06\\
Lup VI &  45 &  0 &  1 &  150$\pm$20 &  1.0 &  6.7$\pm$1.8 &  454.9$\pm$141.4 &  67.5$\pm$10 &  11.2$\pm$6.1 &  1.67$\pm$1.00\\
Lup V &  43 &  0 &  0 &  150$\pm$20 &  1.7 &  11.7$\pm$3.1 &  704.7$\pm$223.5 &  60.5$\pm$10 &  10.8$\pm$5.8 &  0.92$\pm$0.55\\
Mus &  12 &  1 &  0 &  160$\pm$20 &  0.9 &  6.8$\pm$1.7 &  335.1$\pm$109.1 &  49.1$\pm$10 &  3.0$\pm$1.6 &  0.44$\pm$0.26\\
Sco &  10 &  2 &  1 &  130$\pm$15 &  1.4 &  7.3$\pm$1.7 &  620.6$\pm$17.4 &  85.2$\pm$22 &  2.5$\pm$1.3 &  0.34$\pm$0.20\\
Ser-Aqu &  1440 &  146 &  96 &  260$\pm$10 &  8.7 &  179.5$\pm$13.8 &  24441.3$\pm$3025.2 &  136.2$\pm$13 &  360.0$\pm$193.9 &  2.01$\pm$1.09\\\hline
Cloud Averages & 156.1$\pm$71.5 & 20.5$\pm$7.9 & 13.4$\pm$5.2 &274.6$\pm$53.3 & 1.8$\pm$0.5 & 32.4$\pm$9.6 & 2965.1$\pm$1204.9 &79.3$\pm$5.8 & 39.0$\pm$17.9 & 1.2$\pm$0.2\\
Cloud Total & 3122.0 & 409.0 & 268.0 & - & 36.0 & 648.3 & 59302.7 & 91.5 & 780.5 & 1.2 \\\hline
Data from Literature:\\
 & & & & & & &  & &\\
Taurus$^{\bf I}$  &148&-&- & 137&44&252 &27207&108& 37& 0.147\\\hline
\enddata
\tablecomments{  Columns are :
(1) Cloud name.;
(2) Total number of YSOs at all $A_{V}$.;
(3) Number of Class I objects at all $A_{V}$ .;
(4) Number of Flat SED objects  at all $A_{V}$.;
(5) Distances to each cloud.;
(6) Solid angle.;
(7) Area (pc$^{2}$).;
(8) Mass ($M_{\sun}$).;
(9) Surface gas density  ($M_{\sun}$ pc$^{-2}$).;
(10) Star formation rate ($M_{\sun}$ Myr$^{-1}$).;
(11) Star formation rate density ($M_{\sun}$ yr$^{-1}$ kpc$^{-2}$).;
(${\bf I}$)  Total $A_{V}$ mass from \citet{pineda10} and YSO data from \citet{rebull10}.
}
\end{deluxetable}
\end{landscape}

\begin{landscape}
\begin{deluxetable}{lcccrrrrrrrr}
\tabletypesize{\scriptsize}
\tablewidth{0pt}
\tablecaption{ Measured Quantities for Clouds in $A_{V}$ Contours}
\tablehead{
\colhead{Cloud} &
\colhead{$N_{\rm YSOs, I}$}&
\colhead{$N_{\rm YSOs, F}$}&
\colhead{Contour} &
\colhead{$\Omega$}&
\colhead{$A_{\rm con}$} &
\colhead{$M_{\rm con}$}&
\colhead{$\Sigma_{\rm gas, con}$}&
\colhead{SFR, I}&
\colhead{SFR, F}&
\colhead{$\Sigma_{\rm SFR, I}$} &
\colhead{$\Sigma_{\rm SFR, F}$}\\
\colhead{}&
\colhead{}&
\colhead{}&
\colhead{levels$^{\bf I}$}&
\colhead{(deg$^{2}$)}&
\colhead{(pc$^{2}$)}&
\colhead{($M_{\sun}$)}&
\colhead{($M_{\sun}$}&
\colhead{($M_{\sun}$}&
\colhead{($M_{\sun}$}&
\colhead{($M_{\sun}$ yr$^{-1}$} &
\colhead{($M_{\sun}$ yr$^{-1}$}\\
\colhead{}&
\colhead{}&
\colhead{}&
\colhead{(mag)}&
\colhead{}&
\colhead{}&
\colhead{}&
\colhead{pc$^{-2}$)}&
\colhead{Myr$^{-1}$)}&
\colhead{Myr$^{-1}$)}&
\colhead{kpc$^{-2}$)} &
\colhead{kpc$^{-2}$)}\\
\colhead{(1)} &
\colhead{(2)} &
\colhead{(3)} &
\colhead{(4)} &
\colhead{(5)} &
\colhead{(6)} &
\colhead{(7)} &
\colhead{(8)} &
\colhead{(9)} &
\colhead{(10)} &
\colhead{(11)} &
\colhead{(12)}\\
}
\startdata
Cha II &  0 &  0 &  5.2 &  0.8 &  7.78$\pm$1.57 &  416.9$\pm$223.1 &  53.6$\pm$26.6 &  0.9$\pm$1.0* &  1.4$\pm$1.0* &  0.1$\pm$0.1* &  0.2$\pm$0.1*\\
Cha II &  0 &  1 &  8.2 &  0.2 &  1.76$\pm$0.36 &  162.6$\pm$58.7 &  92.3$\pm$27.6 &  0.9$\pm$1.0* &  1.4$\pm$1.0 &  0.5$\pm$0.6* &  0.8$\pm$0.6\\
Cha II &  0 &  1 &  11.8 &  0.03 &  0.30$\pm$0.06 &  43.9$\pm$12.7 &  147.0$\pm$30.4 &  0.9$\pm$1.0* &  1.4$\pm$1.0 &  3.0$\pm$3.3* &  4.7$\pm$3.3\\
Cha II &  1 &  0 &  16.0 &  0.01 &  0.07$\pm$0.01 &  14.0$\pm$3.8 &  193.9$\pm$34.7 &  0.9$\pm$1.0 &  1.4$\pm$1.0* &  12.6$\pm$13.8 &  19.2$\pm$13.8*\\
Lup I &  1 &  0 &  6.0 &  1.2 &  8.05$\pm$2.15 &  417.5$\pm$273.2 &  51.9$\pm$31.0 &  0.9$\pm$1.0 &  1.4$\pm$1.0* &  0.1$\pm$0.1 &  0.2$\pm$0.1*\\
Lup I &  0 &  0 &  10.0 &  0.1 &  0.71$\pm$0.19 &  77.8$\pm$31.0 &  109.3$\pm$32.4 &  0.9$\pm$1.0* &  1.4$\pm$1.0* &  1.3$\pm$1.4* &  2.0$\pm$1.4*\\
Lup I &  0 &  0 &  16.0 &  0.01 &  0.09$\pm$0.02 &  17.2$\pm$5.8 &  184.1$\pm$39.0 &  0.9$\pm$1.0* &  1.4$\pm$1.0* &  9.8$\pm$10.7* &  14.9$\pm$10.7*\\
Lup III &  1 &  2 &  8.0 &  1.2 &  14.89$\pm$2.98 &  815.8$\pm$490.4 &  54.8$\pm$31.1 &  0.9$\pm$1.0 &  2.8$\pm$1.5 &  0.1$\pm$0.1 &  0.2$\pm$1.4\\
Lup III &  0 &  1 &  14.0 &  0.03 &  0.42$\pm$0.08 &  65.0$\pm$20.3 &  153.1$\pm$36.6 &  0.9$\pm$1.0* &  1.4$\pm$1.0 &  2.1$\pm$2.4* &  3.3$\pm$2.4\\
Lup III &  1 &  0 &  20.0 &  0.01 &  0.13$\pm$0.03 &  31.3$\pm$8.7 &  249.0$\pm$48.0 &  0.9$\pm$1.0 &  1.4$\pm$1.0* &  7.2$\pm$8.0 &  11.1$\pm$8.0*\\
Lup IV &  0 &  0 &  8.0 &  0.3 &  2.26$\pm$0.60 &  139.0$\pm$79.8 &  61.4$\pm$31.2 &  0.9$\pm$1.0* &  1.4$\pm$1.0* &  0.4$\pm$0.4* &  0.6$\pm$0.4*\\
Lup IV &  0 &  0 &  14.0 &  0.02 &  0.17$\pm$0.04 &  26.0$\pm$9.2 &  156.6$\pm$36.5 &  0.9$\pm$1.0* &  1.4$\pm$1.0* &  5.5$\pm$6.0* &  8.4$\pm$6.0*\\
Lup IV &  0 &  0 &  23.0 &  0.01 &  0.09$\pm$0.02 &  24.3$\pm$7.8 &  266.8$\pm$46.9 &  0.9$\pm$1.0* &  1.4$\pm$1.0* &  10.0$\pm$11.0* &  15.3$\pm$11.0*\\
Oph &  0 &  2 &  10.5 &  5.6 &  26.61$\pm$10.64 &  2323.2$\pm$1421.2 &  87.3$\pm$40.4 &  0.9$\pm$1.0* &  2.8$\pm$1.5 &  0.03$\pm$0.04* &  0.1$\pm$1.4\\
Oph &  1 &  3 &  18.0 &  0.4 &  1.86$\pm$0.75 &  368.1$\pm$170.7 &  197.4$\pm$46.3 &  0.9$\pm$1.0 &  4.2$\pm$2.2 &  0.5$\pm$0.5 &  2.2$\pm$1.7\\
Oph &  5 &  5 &  25.5 &  0.1 &  0.57$\pm$0.23 &  182.3$\pm$80.1 &  319.6$\pm$58.1 &  4.5$\pm$2.5 &  6.9$\pm$3.7 &  8.0$\pm$5.4 &  12.2$\pm$8.2\\
Oph &  9 &  14 &  33.0 &  0.1 &  0.34$\pm$0.14 &  147.5$\pm$63.8 &  436.7$\pm$72.0 &  8.2$\pm$4.5 &  19.4$\pm$10.5 &  24.2$\pm$16.4 &  57.6$\pm$38.6\\
Oph &  10 &  12 &  41.0 &  0.04 &  0.17$\pm$0.07 &  94.3$\pm$40.8 &  541.8$\pm$89.8 &  9.1$\pm$5.0 &  16.7$\pm$9.0 &  52.2$\pm$35.4 &  95.8$\pm$64.2\\
Per &  3 &  0 &  6.5 &  2.7 &  52.28$\pm$20.91 &  3504.2$\pm$2158.9 &  67.0$\pm$31.4 &  2.7$\pm$1.5 &  1.4$\pm$1.0* &  0.1$\pm$1.7 &  0.03$\pm$0.02*\\
Per &  15 &  4 &  11.0 &  0.8 &  15.39$\pm$6.16 &  1880.9$\pm$910.5 &  122.2$\pm$33.3 &  13.6$\pm$7.5 &  5.6$\pm$3.0 &  0.9$\pm$3.9 &  0.4$\pm$2.0\\
Per &  18 &  14 &  15.5 &  0.2 &  3.79$\pm$1.52 &  734.4$\pm$331.1 &  193.6$\pm$40.3 &  16.4$\pm$9.0 &  19.4$\pm$10.5 &  4.3$\pm$4.2 &  5.1$\pm$3.7\\
Per &  28 &  11 &  20.0 &  0.1 &  1.49$\pm$0.60 &  388.8$\pm$170.4 &  260.3$\pm$46.7 &  25.5$\pm$13.9 &  15.3$\pm$8.2 &  17.0$\pm$11.6 &  10.2$\pm$6.9\\
Per &  8 &  1 &  24.5 &  0.01 &  0.22$\pm$0.09 &  69.8$\pm$30.2 &  316.9$\pm$52.8 &  7.3$\pm$4.0 &  1.4$\pm$1.0 &  33.0$\pm$22.4 &  6.3$\pm$4.5\\
Per &  0 &  0 &  30.0 &  0.001 &  0.02$\pm$0.01 &  7.2$\pm$3.2 &  403.0$\pm$81.0 &  0.9$\pm$1.0* &  1.4$\pm$1.0* &  50.9$\pm$56.0* &  77.8$\pm$56.0*\\
Ser &  2 &  1 &  10.2 &  0.6 &  13.34$\pm$1.03 &  1588.5$\pm$479.4 &  119.1$\pm$34.7 &  1.8$\pm$1.0 &  1.4$\pm$1.0 &  0.1$\pm$1.4 &  0.1$\pm$0.1\\
Ser &  2 &  2 &  14.5 &  0.1 &  2.54$\pm$0.20 &  456.7$\pm$104.0 &  179.9$\pm$38.5 &  1.8$\pm$1.0 &  2.8$\pm$1.5 &  0.7$\pm$1.4 &  1.1$\pm$1.4\\
Ser &  7 &  5 &  18.8 &  0.04 &  0.91$\pm$0.07 &  221.5$\pm$44.3 &  243.2$\pm$44.9 &  6.4$\pm$3.5 &  6.9$\pm$3.7 &  7.0$\pm$3.9 &  7.6$\pm$4.1\\
Ser &  20 &  9 &  23.0 &  0.01 &  0.23$\pm$0.02 &  70.1$\pm$13.3 &  306.6$\pm$53.3 &  18.2$\pm$9.9 &  12.5$\pm$6.7 &  79.6$\pm$44.0 &  54.7$\pm$29.8\\
Aur &  19 &  6 &  8.8 &  1.8 &  48.05$\pm$9.61 &  4293.9$\pm$1001.2 &  89.4$\pm$10.7 &  17.3$\pm$9.4 &  8.3$\pm$4.5 &  0.4$\pm$4.4 &  0.2$\pm$2.4\\
Aur &  20 &  15 &  15.5 &  0.1 &  1.77$\pm$0.35 &  274.0$\pm$60.1 &  154.9$\pm$13.9 &  18.2$\pm$9.9 &  20.8$\pm$11.2 &  10.3$\pm$6.0 &  11.8$\pm$6.8\\
Aur &  2 &  1 &  22.2 &  0.01 &  0.18$\pm$0.04 &  49.7$\pm$11.8 &  282.8$\pm$35.8 &  1.8$\pm$1.0 &  1.4$\pm$1.0 &  10.4$\pm$6.0 &  7.9$\pm$5.7\\
AurN &  0 &  0 &  5.2 &  0.01 &  0.46$\pm$0.09 &  30.9$\pm$7.8 &  67.4$\pm$10.4 &  0.9$\pm$1.0* &  1.4$\pm$1.0* &  2.0$\pm$2.2* &  3.0$\pm$2.2*\\
AurN &  1 &  0 &  8.3 &  0.1 &  1.95$\pm$0.39 &  193.0$\pm$44.2 &  98.8$\pm$11.0 &  0.9$\pm$1.0 &  1.4$\pm$1.0* &  0.5$\pm$0.5 &  0.7$\pm$0.5*\\
Cep &  13 &  3 &  7.5 &  1.3 &  35.77$\pm$35.77 &  2327.7$\pm$597.5 &  331.9$\pm$52.1 &  11.8$\pm$7.5 &  4.2$\pm$3.0 &  0.3$\pm$0.2 &  0.1$\pm$0.1\\
Cep &  13 &  3 &  13.0 &  0.1 &  2.21$\pm$2.21 &  282.6$\pm$63.3 &  507.2$\pm$51.3 &  11.8$\pm$7.0 &  4.2$\pm$3.0 &  5.4$\pm$3.3 &  1.9$\pm$1.4\\
Cha I &  3 &  6 &  8.0 &  0.6 &  7.44$\pm$1.49 &  544.1$\pm$136.0 &  73.1$\pm$11.0 &  2.7$\pm$1.5 &  8.3$\pm$4.5 &  0.4$\pm$1.7 &  1.1$\pm$2.4\\
Cha I &  7 &  5 &  14.0 &  0.1 &  1.75$\pm$0.35 &  256.0$\pm$57.5 &  146.6$\pm$15.0 &  6.4$\pm$3.5 &  6.9$\pm$3.7 &  3.6$\pm$2.6 &  4.0$\pm$2.3\\
Cha I &  0 &  1 &  21.0 &  0.02 &  0.22$\pm$0.04 &  56.0$\pm$13.1 &  253.6$\pm$30.9 &  0.9$\pm$1.0* &  1.4$\pm$1.0 &  4.1$\pm$4.5* &  6.3$\pm$4.5\\
Cha III &  1 &  0 &  5.0 &  2.2 &  26.85$\pm$5.37 &  1228.5$\pm$364.1 &  45.8$\pm$10.0 &  0.9$\pm$1.0 &  1.4$\pm$1.0* &  0.03$\pm$0.04 &  0.05$\pm$0.04*\\
Cha III &  0 &  0 &  8.0 &  0.1 &  1.14$\pm$0.23 &  97.5$\pm$23.0 &  85.2$\pm$10.6 &  0.9$\pm$1.0* &  1.4$\pm$1.0* &  0.8$\pm$0.9* &  1.2$\pm$0.9*\\
CrA &  2 &  1 &  9.3 &  0.5 &  2.45$\pm$0.94 &  162.8$\pm$68.1 &  66.5$\pm$11.0 &  1.8$\pm$1.0 &  1.4$\pm$1.0 &  0.7$\pm$1.4 &  0.6$\pm$0.4\\
CrA &  1 &  2 &  16.7 &  0.09 &  0.47$\pm$0.18 &  87.2$\pm$34.7 &  183.5$\pm$18.2 &  0.9$\pm$1.0 &  2.8$\pm$1.5 &  1.9$\pm$2.1 &  5.8$\pm$3.9\\
CrA &  4 &  0 &  24.0 &  0.02 &  0.10$\pm$0.04 &  29.1$\pm$11.6 &  289.8$\pm$28.4 &  3.6$\pm$2.0 &  1.4$\pm$1.0* &  36.2$\pm$24.2 &  13.8$\pm$9.9*\\
IC5146E &  0 &  0 &  4.7 &  0.2 &  50.60$\pm$8.52 &  2424.3$\pm$673.2 &  47.9$\pm$10.6 &  0.9$\pm$1.0* &  1.4$\pm$1.0* &  0.02$\pm$0.02* &  0.03$\pm$0.02*\\
IC5146E &  11 &  6 &  7.4 &  0.04 &  10.82$\pm$1.82 &  940.8$\pm$204.1 &  86.9$\pm$11.9 &  10.0$\pm$5.5 &  8.3$\pm$4.5 &  0.9$\pm$3.3 &  0.8$\pm$2.4\\
IC5146NW &  7 &  0 &  5.5 &  0.3 &  73.75$\pm$12.42 &  3831.8$\pm$989.0 &  52.0$\pm$10.2 &  6.4$\pm$3.5 &  1.4$\pm$1.0* &  0.1$\pm$2.6 &  0.02$\pm$0.01*\\
IC5146NW &  8 &  3 &  9.0 &  0.05 &  13.83$\pm$2.33 &  1346.3$\pm$275.4 &  97.3$\pm$11.3 &  7.3$\pm$4.0 &  4.2$\pm$2.2 &  0.5$\pm$2.8 &  0.3$\pm$1.7\\
Lup V &  0 &  0 &  4.5 &  1.3 &  9.08$\pm$2.42 &  514.5$\pm$166.2 &  56.7$\pm$10.3 &  0.9$\pm$1.0* &  1.4$\pm$1.0* &  0.1$\pm$0.1* &  0.2$\pm$0.1*\\
Lup V &  0 &  0 &  7.0 &  0.4 &  2.57$\pm$0.69 &  190.0$\pm$57.5 &  73.9$\pm$10.6 &  0.9$\pm$1.0* &  1.4$\pm$1.0* &  0.4$\pm$0.4* &  0.5$\pm$0.4*\\
Lup VI &  0 &  0 &  4.5 &  0.6 &  3.83$\pm$1.02 &  237.4$\pm$75.4 &  62.0$\pm$10.7 &  0.9$\pm$1.0* &  1.4$\pm$1.0* &  0.2$\pm$0.3* &  0.4$\pm$0.3*\\
Lup VI &  0 &  0 &  9.0 &  0.4 &  2.90$\pm$0.77 &  216.4$\pm$65.8 &  74.6$\pm$10.9 &  0.9$\pm$1.0* &  1.4$\pm$1.0* &  0.3$\pm$0.3* &  0.5$\pm$0.3*\\
Mus &  1 &  0 &  4.5 &  0.8 &  5.86$\pm$1.47 &  259.2$\pm$88.0 &  44.2$\pm$10.2 &  0.9$\pm$1.0 &  1.4$\pm$1.0* &  0.2$\pm$0.2 &  0.2$\pm$0.2*\\
Mus &  0 &  0 &  7.0 &  0.1 &  0.95$\pm$0.24 &  75.4$\pm$21.4 &  79.1$\pm$10.7 &  0.9$\pm$1.0* &  1.4$\pm$1.0* &  1.0$\pm$1.0* &  1.5$\pm$1.0*\\
Sco &  1 &  1 &  7.5 &  1.2 &  6.35$\pm$6.35 &  484.7$\pm$130.6 &  456.4$\pm$63.9 &  0.9$\pm$1.0 &  1.4$\pm$1.0 & 0.1$\pm$0.2 &  0.2$\pm$0.2\\
Sco &  1 &  0 &  17.0 &  0.2 &  0.94$\pm$0.94 &  135.8$\pm$34.3 &  567.1$\pm$57.6 &  0.9$\pm$1.0 &  1.4$\pm$1.0* &  1.0$\pm$1.1 &  1.5$\pm$1.1*\\
Ser-Aqu &  0 &  0 &  7.0 &  3.0 &  61.01$\pm$4.69 &  5509.9$\pm$782.3 &  90.3$\pm$10.8 &  0.9$\pm$1.0* &  1.4$\pm$1.0* &  0.02$\pm$0.02* &  0.02$\pm$0.02*\\
Ser-Aqu &  9 &  4 &  12.0 &  4.6 &  95.23$\pm$7.33 &  12818.2$\pm$1540.2 &  134.6$\pm$12.4 &  8.2$\pm$4.5 &  5.6$\pm$3.0 &  0.1$\pm$3.0 &  0.1$\pm$2.0\\
Ser-Aqu &  16 &  12 &  17.0 &  0.6 &  12.57$\pm$0.97 &  2658.5$\pm$298.5 &  211.4$\pm$17.3 &  14.5$\pm$8.0 &  16.7$\pm$9.0 &  1.2$\pm$4.0 &  1.3$\pm$3.5\\
Ser-Aqu &  31 &  20 &  22.0 &  0.3 &  6.66$\pm$0.51 &  1929.6$\pm$219.8 &  289.6$\pm$24.3 &  28.2$\pm$15.4 &  27.8$\pm$15.0 &  4.2$\pm$5.6 &  4.2$\pm$4.5\\
Ser-Aqu &  31 &  25 &  27.0 &  0.1 &  2.88$\pm$0.22 &  1036.8$\pm$125.3 &  359.6$\pm$33.5 &  28.2$\pm$15.4 &  34.7$\pm$18.7 &  9.8$\pm$5.6 &  12.0$\pm$6.6\\
Ser-Aqu &  50 &  24 &  33.0 &  0.1 &  1.10$\pm$0.08 &  488.4$\pm$65.1 &  442.5$\pm$48.2 &  45.5$\pm$24.9 &  33.3$\pm$18.0 &  41.2$\pm$22.8 &  30.2$\pm$16.4\\\hline
Contour Averages & 6.6$\pm$1.2 & 4.1$\pm$0.7 & -999.0$\pm$-999.0 &0.6$\pm$0.1 & 10.6$\pm$2.5 & 972.1$\pm$248.0 & 188.1$\pm$17.7 &6.0$\pm$1.1 & 5.7$\pm$1.0 & 7.7$\pm$2.0 & 8.6$\pm$2.3 \\
\enddata
\tablecomments{  Columns are :
(1) Cloud name.;
(2) Number of Class I YSOs in contour level.;
(3) Number of Flat SED YSOs in contour level.;
(4) $A_{V}$ contour level in mag at which mass measurement was made. The contour levels start at $A_{V}$=2 or the cloud completeness
 limit and increase in even intervals to the listed contour level.;
(5) Solid angle.;
(6) Area in contour level (pc$^{-2}$).;
(7) Mass in contour level ($M_{\sun}$).;
(8) Surface gas density  in contour level  ($M_{\sun}$ pc$^{-2}$).;
(9) Star formation rate in contour level  ($M_{\sun}$ Myr$^{-1}$) for Class I YSOs. Asterisks denote that measurement is an upper limit.;
(10) Star formation rate in contour level  ($M_{\sun}$ Myr$^{-1}$) for Flat SED YSOs. Asterisks denote that measurement is an upper limit.;
(11) Star formation rate density in contour level  ($M_{\sun}$ yr$^{-1}$ kpc$^{-2}$) for  Class I  YSOs. Asterisks denote that measurement is an upper limit.;
(12) Star formation rate density in contour level  ($M_{\sun}$ (yr$^{-1}$ kpc$^{-2}$) for Flat SED YSOs. Asterisks denote that measurement is an upper limit.;
(${\bf I}$) Contour levels start at $A_{V}=2$ for all clouds except for Serpens and Ophiuchus which are covered by the c2d survey completely down to $A_{\rm V}$ = 6 and 3
as discussed in Section~\ref{sdent}.
}
\label{tcont}
\end{deluxetable}
\end{landscape}

\begin{deluxetable}{lrrlrrrlrl}
\tabletypesize{\scriptsize}
\tablewidth{0pt}
\tablecaption{Properties of Suspicious YSOs and MISFITS}
\tablehead{
\colhead{Cloud} &
\colhead{RA} &
\colhead{DEC} &
\colhead{Classification} &
\colhead{$\alpha$} &
\colhead{$\int T_{\rm MB}\ {\rm d}V$} &
\colhead{$T_{\rm MB}$} &
\colhead{SED} &
\colhead{$A_{V}$}  &
\colhead{comments}\\
\colhead{} &
\colhead{J2000} &
\colhead{J2000} &
\colhead{} &
\colhead{} &
\colhead{(K km s$^{-1}$)} &
\colhead{(K)} &
\colhead{class} &
\colhead{(mag)} &
\colhead{} \\
\colhead{(1)} &
\colhead{(2)} &
\colhead{(3)} &
\colhead{(4)} &
\colhead{(5)} &
\colhead{(6)} &
\colhead{(7)} &
\colhead{(8)} &
\colhead{(9)}  &
\colhead{(10)}\\
}
\startdata
Aur      &04:18:21.27  &$+$38:01:35.88  &YSOc			   &$-$0.09   	&                    &$<$0.23             &Flat 	&4   &  \\
Aur      &04:19:44.67  &$+$38:11:21.98  &YSOc\_star$+$dust(IR1)   &$-$0.07   	&                    &$<$0.23             &Flat		&7   &  \\
Aur      &04:29:40.02  &$+$35:21:08.95  &YSOc\_star$+$dust(IR1)   &0.51   	&                    &$<$0.31             &I 		&8   &  \\
Aur      &04:30:14.96  &$+$36:00:08.53  &YSOc\_red		   &1.77  	&  0.42$\pm$0.09     &0.69$\pm$0.15       &I 		&8   &  \\
Aur     &04:30:23.83  &$+$35:21:12.35  &YSOc\_red                &0.61   	&                    &$<$0.30             &I            &8   &  \\
Aur     &04:30:41.17  &$+$35:29:41.08  &YSOc\_red		   &1.49   	&  0.76$\pm$0.07     &1.49$\pm$0.14       &I  		&7   &  self--reversed \\
Aur     &04:30:44.23  &$+$35:59:51.16  &YSOc			   &1.08   	&  0.76$\pm$0.15     &0.78$\pm$0.15       &I  		&8   &  \\
Aur     &04:30:48.52  &$+$35:37:53.76  &YSOc\_red                &1.46   	&  1.01$\pm$0.13     &1.15$\pm$0.15       &I            &7   & self--reversed \\
Aur     &04:30:56.62  &$+$35:30:04.55  &YSOc\_red		   &2.35        &  0.49$\pm$0.07     &1.00$\pm$0.15       &I   		&7   &  \\
Cep &   22:29:33.35 &  $+$75:13:16.01 &   YSOc\_red &         0.20 &               &  $<$0.42   &   Flat &   6 & \\
Cep &   22:35:00.82 &  $+$75:15:36.42 &   YSOc\_star$+$dust(IR2) &        -0.29 &               &  $<$0.46   &   Flat &   6 & \\
Cep &   22:35:14.09 &  $+$75:15:02.61 &   YSOc\_red &         0.36 &               &  $<$0.42   &   I &   6 & \\
Cep &   21:01:36.07 &  $+$68:08:22.54 &   YSOc &        -0.21 &               &  $<$0.42   &   I &   5 & \\
Cep &   21:01:43.89 &  $+$68:14:03.31 &   YSOc\_red &         0.14 &               &  $<$0.40   &   I &   6 & \\
Cep &   21:02:14.06 &  $+$68:07:30.80 &   YSOc\_red &         0.49 &               &  $<$0.40   &   I &   6 & \\
Cep &   21:02:21.22 &  $+$67:54:20.28 &   YSOc\_red &         0.68 &         0.89$\pm$      0.18 &         1.74$\pm$      0.34 &   I &   9 &  double peak\\
Cep &    21:02:21.22 &  $+$67:54:20.28 &   YSOc\_red &         0.68             &         0.72$\pm$      0.11 &	 2.33$\pm$      0.34 &   I &   9 & double peak\\
Cep &   21:02:21.36 &  $+$68:04:36.11 &   YSOc\_PAH$-$em &         0.52 &               &  $<$0.42   &   I &   5 & \\
Cep &   21:02:59.46 &  $+$68:06:32.24 &   YSOc\_red &         0.65 &               &  $<$0.42   &   I &   5 & \\
IC5146E &   21:52:46.58 &  $+$47:12:49.32 &   YSOc\_star$+$dust(IR2) &        -0.19 &               &  $<$0.34   &   Flat &   5 & \\
IC5146E &   21:53:36.24 &  $+$47:10:27.84 &   YSOc\_star$+$dust(IR1) &        -0.12 &               &  $<$0.30   &   Flat &   6 & \\
IC5146E &   21:54:18.76 &  $+$47:12:09.73 &   YSOc\_star$+$dust(IR2) &        -0.23 &               &  $<$0.26   &   Flat &   4 & \\
IC5146E &   21:52:14.36 &  $+$47:14:54.60 &   YSOc\_star$+$dust(IR2) &         0.67 &               &  $<$0.28   &   I &   4 & \\
IC5146E &   21:52:37.78 &  $+$47:14:38.40 &   YSOc\_star+dust(IR1) &         0.64 &         1.59$\pm$      0.25 &         1.80$\pm$      0.28 &   I &   5 & \\
IC5146E &   21:53:06.94 &  $+$47:14:34.80 &   YSOc &         0.34 &         0.71$\pm$      0.27 &         0.66$\pm$      0.25 &   I &   5 & \\
IC5146E &   21:53:55.70 &  $+$47:20:30.13 &   YSOc\_PAH$-$em &         1.59 &               &  $<$0.34   &   I &   4 & \\
IC5146NW &   21:45:31.22 &  $+$47:36:21.24 &   YSOc &         0.13 &         0.44$\pm$      0.19 &         0.61$\pm$      0.26 &   Flat &   5 & \\
IC5146NW &   21:44:43.08 &  $+$47:46:43.68 &   YSOc\_red &         0.63 &         1.92$\pm$      0.11 &         4.18$\pm$      0.23 &   I &   4 & \\
IC5146NW &   21:44:48.31 &  $+$47:44:59.64 &   YSOc\_red &         1.83 &         0.65$\pm$      0.05 &         3.11$\pm$      0.23 &   I &   5 & \\
IC5146NW &   21:44:53.98 &  $+$47:45:43.56 &   YSOc\_star$+$dust(IR1) &         0.76 &         0.70$\pm$      0.11 &         1.75$\pm$      0.28 &   I &   4 & \\
IC5146NW &   21:45:02.64 &  $+$47:33:07.56 &   YSOc\_red &         1.18 &         0.77$\pm$      0.22 &         1.03$\pm$      0.30 &   I &   4 & \\
IC5146NW &   21:45:08.31 &  $+$47:33:05.77 &   YSOc\_red &         0.74 &         3.72$\pm$      0.53 &         2.07$\pm$      0.30 &   I &   4 & \\
IC5146NW &   21:45:27.86 &  $+$47:45:50.40 &   YSOc\_star$+$dust(IR4) &         0.42 &               &  $<$0.36   &   I &   3 & \\
IC5146NW &   21:47:06.02 &  $+$47:39:39.24 &   YSOc\_red &         0.43 &         0.40$\pm$      0.14 &         0.70$\pm$      0.25 &   I &   5 & \\
Lup I     &15:38:48.35  &$-$34:40:38.24  &YSOc\_PAH$-$em           &0.31   	&                    &$<$0.42             &I    	&3   &  \\
Lup I     &15:43:02.29  &$-$34:44:06.22  &YSOc\_star$+$dust(IR1)   &0.14   	&                    &$<$0.38             &Flat 	&$<$2 & \\
Lup III &   16:07:03.85 &  $-$39:11:11.59 &   YSOc\_star$+$dust(IR1) &        -0.14 &               &  $<$0.30   &   Flat &  $<$2 & \\
Lup III &   16:07:08.57 &  $-$39:14:07.75 &   YSOc &        -0.01 &               &  $<$0.32   &   Flat &  $<$2 & \\
Lup III &   16:07:54.73 &  $-$39:15:44.49 &   YSOc\_red &        -0.15 &               &  $<$0.28   &   Flat &   2 & \\
Lup IV &   16:02:21.61 &  $-$41:40:53.70 &   YSOc\_PAH-em &         0.56 &             &  $<$0.36 &   I &   4 & \\
Lup VI &   16:24:51.78 &  $-$39:56:32.66 &   YSOc &         0.22 &               &  $<$0.48   &   Flat &   8 & \\
Oph    &16:21:38.72  &$-$22:53:28.26  &YSOc\_star$+$dust(IR1)   &0.99   	&                    &$<$0.35             &I    	&$<$3 & \\
Oph    &16:23:40.00  &$-$23:33:37.36  &YSOc                     &0.01   	&                    &$<$0.37             &Flat 	&3    & \\
Oph   &16:44:24.27  &$-$24:01:24.56  &YSOc\_PAH$-$em           &0.27   	&                    &$<$0.35             &Flat 	&$<$3 & \\
Oph   &16:45:26.65  &$-$24:03:05.41  &YSOc\_red                &0.37   	&                    &$<$0.36             &I    	&$<$3 &  \\
Oph &   16:21:45.13 &  $-$23:42:31.63 &   YSOc\_star$+$dust(IR1) &         0.30 &               &  $<$0.36   &   I &   9 & \\
Oph &   16:31:31.24 &  $-$24:26:27.87 &   YSOc\_star$+$dust(IR4) &        -0.24 &               &  $<$0.38   &   Flat &  $<$3 & \\
Oph &   16:25:27.56 &  $-$24:36:47.55 &   YSOc\_star$+$dust(IR1) &         0.06 &               &  $<$0.42   &   Flat &   7 & \\
Oph &   16:23:32.22 &  $-$24:25:53.82 &   YSOc\_star$+$dust(IR2) &        -0.04 &         0.36$\pm$      0.17 &         0.66$\pm$      0.31 &   Flat &   4 & \\
Oph &   16:22:20.99 &  $-$23:04:02.35 &   YSOc\_PAH$-$em &         0.17 &               &  $<$0.44   &   Flat &   4 & \\
Oph &   16:23:05.43 &  $-$23:02:56.73 &   YSOc\_star$+$dust(IR2) &        -0.27 &               &  $<$0.46   &   Flat &   4 & \\
Oph &   16:23:06.86 &  $-$22:57:36.61 &   YSOc &        -0.19 &               &  $<$0.44   &   Flat &   5 & \\
Oph &   16:23:40.00 &  $-$23:33:37.36 &   YSOc &         0.01 &               &  $<$0.46   &   Flat &   4 & \\
Per    &03:25:19.52  &$+$30:34:24.16  &YSOc		           &-0.11   	&                    &$<$0.27   	  &Flat	 	&2 &	\\
Per    &03:26:37.47  &$+$30:15:28.08  &YSOc\_red                &0.99     	&  0.20$\pm$0.04     &0.77$\pm$0.14   	  &I 	 	&2 &double peak	\\
Per    &03:26:37.47  &$+$30:15:28.08  &YSOc\_red                &0.99    	&  0.11$\pm$0.03     &0.49$\pm$0.14   	  &I	 	&2 &double peak	\\
Per    &03:28:34.49  &$+$31:00:51.10  &YSOc\_star$+$dust(IR1)   &0.89    	&  0.31$\pm$0.05     &0.95$\pm$0.15   	  &I 	 	&6 &	\\
Per    &03:28:34.94  &$+$30:54:54.55  &YSOc		           &0.01   	&                    &$<$0.32   	  &Flat  	&3 &	\\
Per    &03:29:06.05  &$+$30:30:39.19  &YSOc\_red		   &0.72	&                    &$<$0.30             &I	 	&2 &	\\
Per    &03:29:51.82  &$+$31:39:06.03  &red		           &3.34	&  3.22$\pm$0.20     &2.41$\pm$0.15   	  &I	 	&6 &	\\
Per    &03:30:22.45  &$+$31:32:40.53  &YSOc\_star$+$dust(IR2)   &0.35	&                    &$<$0.34   	  &I	 	&3 &	\\
Per    &03:30:38.21  &$+$30:32:11.93  &YSOc\_star$+$dust(IR2)   &-0.10	&                    &$<$0.29   	  &Flat	 	&4 &	\\
Per    &03:31:14.70  &$+$30:49:55.40  &YSOc\_star$+$dust(IR1)   &-0.09	&                    &$<$0.30   	  &Flat	 	&2 &	\\
Per    &03:31:20.98  &$+$30:45:30.06  &YSOc\_red		   &1.10	&  2.70$\pm$0.17     &2.68$\pm$0.17  	  &I	 	&5 &	\\
Per    &03:44:24.84  &$+$32:13:48.36  &YSOc\_red		   &1.69	&                    &$<$0.31   	  &I	 	&6 &	\\
Per    &03:44:35.34  &$+$32:28:37.18  &YSOc\_red		   &-0.09	&                    &$<$0.32   	  &Flat	 	&3 &	\\
Per    &03:45:13.82  &$+$32:12:10.00  &YSOc\_red		   &0.43	&                    &$<$0.28   	  &I	 	&3 &	\\
Per    &03:47:05.43  &$+$32:43:08.53  &YSOc\_red		   &0.48	&  0.93$\pm$0.09     &1.48$\pm$0.14   	  &I	 	&5 &	\\
Sco &   16:46:58.27 &  $-$09:35:19.76 &   YSOc\_red &         0.66 &               &  $<$0.44   &   I &   12 & \\
Sco &   16:48:28.85 &  $-$14:14:36.45 &   YSOc\_PAH$-$em &         0.48 &               &  $<$0.48   &   I &   5 & \\
Sco &   16:22:04.35 &  $-$19:43:26.76 &   YSOc &         0.02 &               &  $<$0.66   &   Flat &   6 & \\
Ser    &18:28:41.87  &$-$00:03:21.34  &YSOc\_star$+$dust(IR1)   &0.14   	&                    &$<$0.47             & Flat 	&8   &    \\
Ser    &18:28:44.78  &$+$00:51:25.79  &YSOc\_red                &1.05   	&  0.86$\pm$0.23     &0.83$\pm$0.22       &I     	&8   &    \\
Ser    &18:28:44.96  &$+$00:52:03.54  &YSOc\_red                &1.27   	&  1.61$\pm$0.34     &1.36$\pm$0.28       &I    	&8    &   \\
Ser    &18:29:16.18  &$+$00:18:22.71  &YSOc                     &-0.13  	&  1.28$\pm$0.29     &0.98$\pm$0.22       &Flat  	&7     &  \\
Ser    &18:29:40.20  &$+$00:15:13.11  &YSOc\_star$+$dust(IR1)   &0.68   	&                    &$<$0.50             &I    	&$<$6  &  \\
Ser    &18:30:05.26  &$+$00:41:04.58  &red                      &1.24   	&                    &$<$0.44             &I    	&$<$6   & \\
Ser &   18:28:44.01 &  $+$00:53:37.93 &   YSOc\_red &         0.29 &               &  $<$0.42   &   Flat &   7 & \\
Ser &   18:29:27.35 &  $+$00:38:49.75 &   YSOc &         0.24 &               &  $<$0.38   &   Flat &   13 & \\
Ser &   18:29:31.96 &  $+$01:18:42.91 &   YSOc\_star$+$dust(IR1) &0.32 &         0.63$\pm$      0.27 &         0.69$\pm$      0.30 &   I&   11 & \\
Ser--Aqu &   18:13:45.05 &  $-$03:26:02.67 &   YSOc\_star$+$dust(IR1) &         0.39 &               &  $<$0.58   &   I &   6 & \\
Ser--Aqu &   18:27:03.33 &  $-$02:45:33.42 &   YSOc\_red &         0.44 &               &  $<$0.54   &   I &   5 & \\
Ser--Aqu &   18:29:16.80&  $-$01:17:30.68 &   YSOc &         0.78 &               &  $<$0.54   &   I &   10 & \\
Ser--Aqu &   18:30:32.48 &  $-$03:50:01.21 &   YSOc\_star$+$dust(MP1) &         0.37 &               &  $<$0.54   &   I &   9 & \\
Ser--Aqu &   18:33:03.49 &  $-$02:08:42.53 &   YSOc\_PAH$-$em &         1.25 &               &  $<$0.56   &   I &   8 & \\
Ser--Aqu &   18:37:39.24 &  $-$00:25:35.18 &   YSOc\_star$+$dust(IR1) &         0.72 &               &  $<$0.56   &   I &   7 & \\
Ser--Aqu &   18:37:46.90 &  $-$00:01:55.83 &   YSOc\_PAH$-$em &         1.19 &               &  $<$0.52   &   I &   8 & \\
Ser--Aqu &   18:37:52.77 &  $-$00:23:03.10 &   YSOc\_star$+$dust(MP1) &         0.59 &               &  $<$0.50   &   I &   7 & \\
Ser--Aqu &   18:37:55.79 &  $-$00:23:31.59 &   YSOc &         1.14 &  &$<$0.58   &   I  &   8 & \\
Ser--Aqu &   18:05:31.11 &  $-$04:38:09.63 &   YSOc &         0.23 &               &  $<$0.60   &   Flat &   9 & \\
Ser--Aqu &   18:10:28.90 &  $-$02:37:42.79 &   YSOc &        -0.18 &               &  $<$0.62   &   Flat &   6 & \\
Ser--Aqu &   18:26:32.81 &  $-$03:46:27.26 &   YSOc\_red &         0.08 &               &  $<$0.60   &   Flat &   10 & \\
Ser--Aqu &   18:27:24.87 &  $-$03:58:21.15 &   YSOc &        -0.22 &               &  $<$0.60   &   Flat &   10 & \\
Ser--Aqu &   18:28:09.49 &  $-$02:26:31.95 &   YSOc\_star$+$dust(IR2) &        -0.11 &               &  $<$0.58   &   Flat &   5 & \\
Ser--Aqu &   18:29:16.72 &  $-$01:17:36.92 &   YSOc\_star$+$dust(MP1) &        -0.08 &               &  $<$0.62   &   Flat &   10 & \\
Ser--Aqu &   18:30:06.06 &  $-$01:10:19.33 &   YSOc &        -0.15 &               &  $<$0.56   &   Flat &   6 & \\
Ser--Aqu &   18:30:13.01 &  $-$01:25:36.64 &   YSOc\_star$+$dust(IR2) &        -0.23 &               &  $<$0.60   &   Flat &   8 & \\
Ser--Aqu &   18:36:02.64 &  $-$00:02:20.70 &   YSOc\_star$+$dust(MP1) &        -0.28 &               &  $<$0.58   &   Flat &   8 & \\
Ser--Aqu &   18:38:55.77 &  $-$00:23:40.81 &   YSOc &        -0.05 &               &  $<$0.64   &   Flat &   7 & \\
Ser--Aqu &   18:40:12.06 &  $+$00:29:27.74 &   YSOc\_red &       -0.07 &               &  $<$0.50   &   Flat &   8 & \\\hline
\enddata
\tablecomments{  Columns are :
(1) Cloud
(2) Source Right Ascension in J2000 coordinates;
(3) Source Declination in J2000 coordinates;
(4) Source classification (see \citet{evans09});
(5) Spectral Index, extinction corrected values for c2d clouds only;
(6) Integrated main beam \hcop line intensity.;
(7) Main beam \hcop line temperature, upper limits are computed as 2$\sigma_{\rm rms}$;
(8) SED class based on \citet{greene94};
(9) $A_{V}$ in source position. Values that are found outside the
 $A_{V}$ map completeness limit are given as $<$ limit;
(10) Line profile comments.
}
\label{tmis}
\end{deluxetable}

\begin{deluxetable}{lrrrrrr}
\tabletypesize{\scriptsize}
\tablewidth{0pt}
\tablecaption{$A_{V}$, $^{12}$CO, and $^{13}$CO Masses and $\Sigma_{\rm gas}$
 for Per and Oph Clouds}
\tablehead{
\colhead{Cloud} &
\colhead{$M_{\rm cloud,\rm gas}$}&
\colhead{$M_{\rm cloud,^{12}\rm CO}$}&
\colhead{$M_{\rm cloud,^{13}\rm CO}$}&
\colhead{$\Sigma_{\rm cloud, \rm gas}$}&
\colhead{$\Sigma_{\rm cloud,^{12}\rm CO}$}&
\colhead{$\Sigma_{\rm cloud,^{13}\rm CO}$}\\
\colhead{}&
\colhead{($M_{\sun}$)}&
\colhead{($M_{\sun}$)}&
\colhead{($M_{\sun}$)}&
\colhead{($M_{\sun}$}&
\colhead{($M_{\sun}$}&
\colhead{($M_{\sun}$}\\
\colhead{}&
\colhead{}&
\colhead{}&
\colhead{}&
\colhead{pc$^{-2}$)}&
\colhead{pc$^{-2}$)}&
\colhead{pc$^{-2}$)}\\
\colhead{(1)} &
\colhead{(2)} &
\colhead{(3)} &
\colhead{(4)} &
\colhead{(5)} &
\colhead{(6)} &
\colhead{(7)} \\
}
\startdata
Per & 5997$\pm$3387 & 9657$\pm$2416 & 1073$\pm$110 & 82$\pm$33 & 132$\pm$33 & 15$\pm$2\\
Oph & 2270$\pm$1533 & 2596$\pm$659 & 348$\pm$39 & 77$\pm$42 & 88$\pm$22 & 12$\pm$1\\\hline
Data from Literature:\\
Taurus$^{\bf I}$ & 27207 & 16052 &  & 108 & 64 & \\
\hline\hline
\enddata
\tablecomments{
(1) Cloud name.;
(2) Mass from $A_{V}$ map ($M_{\sun}$) where there is positive $^{12}$CO and   $^{13}$CO emission.;
(3) $^{12}$CO Mass ($M_{\sun}$).;
(4) $^{13}$CO Mass ($M_{\sun}$).;
(5) Surface gas density from $A_{V}$ map ($M_{\sun}$ pc$^{-2}$).;
(6) $^{12}$CO Surface gas density  ($M_{\sun}$ pc$^{-2}$).;
(7) $^{13}$CO Surface gas density  ($M_{\sun}$ pc$^{-2}$).;
(${\bf I}$)  Combined $^{12}$ CO and  $^{13}$CO mass from \citet{goldsmith08} and $A_{V}$ mass from \citep{pineda10}.
}
\label{tco}
\end{deluxetable}

\begin{deluxetable}{lrrrrrrr}
\tabletypesize{\scriptsize}
\tablewidth{0pt}
\tablecaption{$A_{V}$, $^{12}$CO, and $^{13}$CO Masses and $\Sigma_{\rm gas}$
 for Per and Oph Clouds in $A_{V}$ Contours}
\tablehead{
\colhead{Cloud} &
\colhead{Contour} &
\colhead{$M_{\rm con,\rm gas}$}&
\colhead{$M_{\rm con, ^{12}\rm CO}$}&
\colhead{$M_{\rm con, ^{13}\rm CO}$}&
\colhead{$\Sigma_{\rm con,\rm gas}$}&
\colhead{$\Sigma_{\rm con,^{12}\rm CO}$}&
\colhead{$\Sigma_{\rm con,^{13}\rm CO}$} \\
\colhead{}&
\colhead{levels}&
\colhead{($M_{\sun}$)}&
\colhead{($M_{\sun}$)}&
\colhead{($M_{\sun}$)}&
\colhead{($M_{\sun}$}&
\colhead{($M_{\sun}$}&
\colhead{($M_{\sun}$} \\
\colhead{}&
\colhead{(mag)}&
\colhead{}&
\colhead{}&
\colhead{}&
\colhead{pc$^{-2}$)}&
\colhead{pc$^{-2}$)}&
\colhead{pc$^{-2}$)} \\
\colhead{(1)} &
\colhead{(2)} &
\colhead{(3)} &
\colhead{(4)} &
\colhead{(5)} &
\colhead{(6)} &
\colhead{(7)} &
\colhead{(8)} \\
}
\startdata
Per & 6.5  & 3144.7$\pm$2068.5  & 6050.8$\pm$1514.5  & 518.0$\pm$54.0  & 60.2$\pm$31 & 115.7$\pm$28  & 9.9$\pm$1\\
 & 11.0  & 1774.0$\pm$875.5  & 2555.9$\pm$639.3  & 349.1$\pm$35.3  & 115.3$\pm$33 & 166.1$\pm$41  & 22.7$\pm$2\\
 & 15.5  & 652.9$\pm$302.6  & 654.0$\pm$163.6  & 120.3$\pm$12.1  & 172.1$\pm$40 & 172.4$\pm$43  & 31.7$\pm$3\\
 & 20.0  & 369.7$\pm$163.5  & 343.7$\pm$86.0  & 76.2$\pm$7.6  & 247.6$\pm$46 & 230.2$\pm$57  & 51.0$\pm$5\\
 & 24.5  & 49.5$\pm$23.0  & 49.4$\pm$12.4  & 8.9$\pm$0.9  & 224.9$\pm$52 & 224.5$\pm$56  & 40.5$\pm$4\\
 & 30.0  & 6.1$\pm$2.8  & 3.3$\pm$0.8  & 0.7$\pm$0.1  & 339.3$\pm$80 & 187.7$\pm$46  & 37.6$\pm$3\\\hline
Oph & 10.5 &    1577.2$\pm$1246.7  & 1982.0$\pm$505  & 183.3$\pm$23  & 76.8$\pm$40 & 74.5$\pm$19  & 6.9$\pm$0\\
 & 18.0 &    349.4$\pm$164.3  & 328.4$\pm$82  & 71.1$\pm$7  & 187.4$\pm$46 & 176.1$\pm$44  & 38.1$\pm$3\\
 & 25.5 &    161.3$\pm$72.5  & 135.9$\pm$34  & 42.6$\pm$4  & 282.8$\pm$58 & 238.3$\pm$59  & 74.7$\pm$7\\
 & 33.0 &   111.5$\pm$50.8  & 97.1$\pm$24  & 32.2$\pm$3  & 330.3$\pm$72 & 287.5$\pm$71  & 95.3$\pm$9\\
 & 41.0 &   70.6$\pm$32.3  & 52.9$\pm$13  & 18.6$\pm$1  & 405.9$\pm$89 & 304.0$\pm$76  & 106.6$\pm$10\\\hline\hline
\enddata
\tablecomments{
(1) Cloud name.;
(2) $A_{V}$ contour level in mag at which mass measurement was made. The contour levels start at $A_{V}$=2 or the cloud completeness
 limit and increase in even intervals to the listed contour level.;
(3) $A_{V}$ Mass ($M_{\sun}$) where there is positive $^{12}$CO and $^{13}$CO emission.;
(4) $^{12}$CO Mass ($M_{\sun}$).;
(5) $^{13}$CO Mass ($M_{\sun}$).;
(6) $A_{V}$ Surface gas density  ($M_{\sun}$ pc$^{-2}$).;
(7) $^{12}$CO Surface gas density  ($M_{\sun}$ pc$^{-2}$).;
(8) $^{13}$CO Surface gas density  ($M_{\sun}$ pc$^{-2}$).
}
\label{tco2}
\end{deluxetable}

\begin{deluxetable}{lcc}
\tabletypesize{\scriptsize}
\tablewidth{0pt}
\tablecaption{Massive Clumps HCN$J$=(1--0)}
\tablehead{
\colhead{Source} &
\colhead{log $\Sigma_{\rm HCN}$} &
\colhead{log $\Sigma_{\rm SFR_{\rm IR}}$} \\
\colhead{} &
\colhead{($M_{\sun}$ pc$^{-2}$)}&
\colhead{($M_{\sun}$ yr$^{-1}$ kpc$^{-2}$)}\\
\colhead{(1)} &
\colhead{(2)} &
\colhead{(3)} \\
}
\startdata
W3(OH) &  3.39$\pm$0.13 &  1.35$\pm$0.12\\
RCW142 &  3.40$\pm$0.14 &  1.08$\pm$0.13\\
W28A2(1) &  3.66$\pm$0.14 &  2.12$\pm$0.13\\
G9.62+0.10 &  3.28$\pm$0.13 &  1.45$\pm$0.14\\
G10.60-0.40 &  3.32$\pm$0.12 &  1.83$\pm$0.12\\
G12.21-0.10 &  2.63$\pm$0.24 &  0.28$\pm$0.23\\
G13.87+0.28 &  2.28$\pm$0.15 &  0.61$\pm$0.13\\
G23.95+0.16 &  2.28$\pm$0.25 &  0.79$\pm$0.21\\
W43S &  2.63$\pm$0.14 &  1.51$\pm$0.14\\
W44 &  2.79$\pm$0.18 &  1.00$\pm$0.16\\
G35.58-0.03 &  3.41$\pm$0.24 &  0.89$\pm$0.22\\
G48.61+0.02 &  1.98$\pm$0.14 &  0.75$\pm$0.13\\
W51M &  3.14$\pm$0.17 &  1.53$\pm$0.17\\
S87 &  2.76$\pm$0.16 &  0.97$\pm$0.12\\
S88B &  2.68$\pm$0.19 &  1.31$\pm$0.15\\
K3-50 &  3.26$\pm$0.12 &  1.75$\pm$0.13\\
ON1 &  2.83$\pm$0.14 &  0.73$\pm$0.13\\
ON2S &  2.76$\pm$0.16 &  1.44$\pm$0.14\\
W75N &  3.13$\pm$0.13 &  1.50$\pm$0.12\\
DR21S &  3.20$\pm$0.13 &  1.75$\pm$0.12\\
W75(OH) &  3.21$\pm$0.13 &  0.62$\pm$0.13\\
CEPA &  3.44$\pm$0.17 &  2.00$\pm$0.13\\
IRAS20126 &  2.98$\pm$0.18 &  1.22$\pm$0.13\\
IRAS20220 &  2.63$\pm$0.21 &  0.29$\pm$0.18\\
IRAS23385 &  2.79$\pm$0.19 &  0.43$\pm$0.15\\\hline
Clump Average & 3.12$\pm$0.16 & 1.44$\pm$0.15\\\hline
\enddata
\tablecomments{  Columns are :
(1) Source name
(2) Surface gas density;
(3) SFR surface gas density.
}
\label{t1msf}
\end{deluxetable}


\end{document}